\documentclass[10pt]{article}

\usepackage{amsmath,amssymb}
\usepackage{graphicx}
\usepackage{caption}
\usepackage{color}
\usepackage{dcolumn}
\usepackage{bm}
\usepackage{float}
\usepackage{hyperref} 
\usepackage{epsfig}
\hypersetup{ colorlinks = true, citecolor = blue, urlcolor = blue}

\definecolor{background-color}{gray}{0.98}

\newcommand{\etal}{{\it et al.}}
\newcommand{\be}{\begin{equation}}
\newcommand{\ee}{\end{equation}}
\newcommand{\bea}{\begin{eqnarray}}
\newcommand{\eea}{\end{eqnarray}}
\newcommand{\Fig}[1]{Fig.\,\ref{#1}}
\newcommand{\Eq}[1]{Eq.\,(\ref{#1})}
\newcommand{\la}{\langle}
\newcommand{\ra}{\rangle}
\newcommand{\average }[1]{ \left\langle #1 \right\rangle }
\newcommand{\trace }{ {\rm Tr} }

\newcommand{\rcite}[1]{Ref.\,\cite{#1}}

\title{The hierarchical and perturbative forms of stochastic Schr\"{o}dinger equations and their applications to carrier dynamics in organic materials}

\author{Yuchen Wang\thanks{State Key Laboratory for Physical Chemistry of Solid Surfaces, Collaborative Innovation Center of Chemistry for Energy Materials, Fujian Provincial Key Lab of Theoretical and Computational Chemistry, and College of Chemistry and Chemical Engineering, Xiamen University, Xiamen 361005, China},
Yaling Ke\footnotemark[1],
Yi Zhao\footnotemark[1] $^,$\footnotemark[2]\thanks{E-mail: yizhao@xmu.edu.cn}}

\date{}
\begin{document}
\maketitle

\begin{center}
\subsubsection*{\small Article Type:}
Advanced Review

\hfill \break
\thanks

\subsubsection*{Abstract}
\begin{flushleft}
A number of non-Markovian stochastic Schr\"odinger equations, ranging from the numerically exact hierarchical form towards a series of perturbative expressions sequentially presented in an ascending degrees of approximations are revisited in this short review, aiming at providing a systematic framework which is capable to connect different kinds of the wavefunction-based approaches for an open system coupled to the harmonic bath. One can optimistically expect the extensive future applications of those non-Markovian stochastic Schr\"odinger equations in large-scale realistic complex systems, benefiting from their favorable scaling with respect to the system size, the stochastic nature which is extremely suitable for parallel computing, and many other distinctive advantages. In addition, we have presented a few examples showing the excitation energy transfer in Fenna-Matthews-Olson complex, a quantitative measure of decoherence timescale of hot exciton, and the study of quantum interference effects upon the singlet fission processes in organic materials, since a deep understanding of both mechanisms is very important to explore the underlying microscopic processes and to provide novel design principles for highly efficient organic photovoltaics.

\end{flushleft}
\end{center}

\clearpage

\renewcommand{\baselinestretch}{1.5}
\normalsize

\clearpage

\section*{\sffamily \Large INTRODUCTION}
With the burst of newly-fabricated materials and rapid developments of experimental techniques, it has been demonstrated by numerous studies that the quantum effects play a fundamental role in ultrafast dynamics, such as the excitation energy transfer in photosynthetic antenna complex\cite{Brixner-Nature2005-p625,Engel-Nature2007-p782,Lee-Science2007-p1462,
Collini-Nature2010-p644,Panitchayangkoon-Proc.Natl.Acad.Sci.USA2010-p12766,Scholes-Nature2017-p647,
Curutchet-Chem.Rev.2016-p,Chenu-Annu.Rev.Phys.Chem.2015-p69} and the carrier dynamics in organic photovoltaics\cite{Banerji-J.Phys.Chem.C2011-p9726,Monahan-Phys.Rev.Lett.2015-p247003,
Gelinas-Science2014-p512,Kaake-J.Phys.Chem.Lett.2013-p2264,Ke-J.Phys.Chem.Lett.2015-p1741,
Bredas-Nat.Mater.2017-p35,Hedley-2016light}.

To elucidate the detailed mechanism behind these processes, it is often necessary to invoke accurate theoretical descriptions, which is challenging due to the large (often nearly innumerous) degrees of freedom in realistic complex systems.
Despite the difficulty, remarkable progress has been made in the past decades.
For the purpose of reducing the number of basis vectors with controllable numerical errors, different brilliant approaches have been developed, the representative examples are the numerical renormalization group\cite{Bulla-Phys.Rev.Lett.2003-p170601,Bulla-Phys.Rev.B2005-p45122}, density matrix renormalization group\cite{White-Phys.Rev.Lett.2004-p76401}, time evolving density matrix using orthogonal polynomials algorithms\cite{Chin-J.Math.Phys.2010-p92109} and multi-layer multi-configuration time-dependent Hartree method (ML-MCTDH)\cite{Wang-J.Chem.Phys.2003-p1289}.
These approaches have been proven to be very efficient in the low temperature regime and have achieved great success in many areas\cite{Duckheim-Phys.Rev.B2005-p134501,Anders-Phys.Rev.Lett.2007-p,Winter-Phys.Rev.Lett.2009-p30601,
Wang-Chem.Phys.2010-p78,Kast-Phys.Rev.B2014-p100301,Kast-Phys.Rev.Lett.2013-p10402}.

Another formalism different from the aforementioned full-space approaches
is based on the reduced system language. In organic materials, for instance, the carriers (usually regarded as the system) can be described by the corresponding reduced density matrix, whereas the vibrational degrees of freedom are mapped into a bosonic bath and are traced over.
This partial trace operation can be done in an elegant way using the path integral technique, leading to the appearance of the famous Feynman-Vernon influence functional in the path integral expression\cite{Feynman-Ann.Phys.1963-p118}, in which all the influences from the bath are included. But this operation also introduces new difficulties since the influence functional is nonlocal both in space and time.
The long time results of the path integral, however, cannot be obtained by directly using the Monte Carlo technique\cite{Herman-J.Chem.Phys.1982-p5150,Egger-ZeitschriftfurPhysikBCondensedMatter1992-p97,
Egger-Phys.Rev.B1994-p15210} due to the notorious sign problem.
By utilizing the well-behaved quasiadiabatic propagators, Makri and coworkers\cite{Makri-Chem.Phys.Lett.1992-p435,Topaler1992multidimensional,Topaler-Chem.Phys.Lett.1993-p285} creatively proposed the quasiadiabatic propagator path integral to overcome this problem.
Whereafter, various numerical optimization techniques, like the discrete value representation\cite{Topaler-Chem.Phys.Lett.1993-p448}, tensor multiplication scheme\cite{Makarov-Chem.Phys.Lett.1994-p482,Makri-J.Chem.Phys.1995-p4600,Makri-J.Chem.Phys.1995-p4611,
Shao-Chem.Phys.2001-p1,Shao-J.Chem.Phys.2002-p507,Makri-J.Math.Phys.1995-p2430} and propagator filtering techniques\cite{Sim-Comput.Phys.Commun.1997-p335} were proposed successively to further lower the cost and extend the applicability of the method.
Another efficient and accurate approach is the hierarchical equations of motion, pioneered by Tanimura and Kubo\cite{Tanimura-J.Phys.Soc.Jpn.1989-p101} for the special case of Drude-Lorentzian spectral densities with the high temperature approximation.
Afterwards, the exact version of hierarchical equations of motion was derived by several contributors via different starting point\cite{Yan-Chem.Phys.Lett.2004-p216,Xu-J.Chem.Phys.2005-p41103,Ishizaki-J.Phys.Soc.Jpn.2005-p3131}.
To date, a lot of efforts have been made to equip hierarchical equations of motion with different powerful techniques, which can be listed as follows: the Pad\'{e} spectrum decomposition\cite{Hu-J.Chem.Phys.2010-p101106}, better truncation schemes\cite{Xu-J.Chem.Phys.2005-p41103,Tanimura-J.Phys.Soc.Jpn.2006-p82001,Schroeder-J.Chem.Phys.2007-p114102}, new decomposition schemes for arbitrary spectral densities\cite{Meier-J.Chem.Phys.1999-p3365} and bath correlation functions\cite{Tang-J.Chem.Phys.2015-p224112,Duan-Phys.Rev.B2017-p214308}, highly efficient filtering algorithm controllable in accuracy\cite{Shi-J.Chem.Phys.2009-p84105,Liu-J.Chem.Phys.2014-p134106}, exponential integrators for time evolution\cite{Wilkins-J.Chem.TheoryComput.2015-p3411}, high-performance computing platform and efficient parallel algorithm\cite{Kreisbeck-J.Chem.TheoryComput.2011-p2166,Kreisbeck-J.Phys.Chem.B2013-p9380,Kreisbeck-J.Phys.Chem.Lett.2012-p2828,
Tsuchimoto-J.Chem.TheoryComput.2015-p3859,Strumpfer-J.Chem.TheoryComput.2012-p2808,Kreisbeck-J.Chem.TheoryComput.2014-p4045}.
Nowadays, hierarchical equations of motion has become a standard approach in the fields of open quantum systems, and has made great contributions to many areas of physics and chemistry\cite{Ishizaki-Proc.Natl.Acad.Sci.U.S.A.2009-p17255,Chen-J.Chem.Phys.2009-p94502,Chen-J.Chem.Phys.2011-p194508,Tanimura-J.Phys.Soc.Jpn.2006-p82001}.
A different strategy from quasiadiabatic propagator path integral and hierarchical equations of motion to treat the influence functional is the direct stochastic unravelling by introducing two correlated Gaussian complex stochastic processes, leading to the stochastic Liouville-von Neumann equation\cite{Stockburger-Phys.Rev.Lett.1998-p2657,Stockburger-J.Chem.Phys.1999-p4983,
Stockburger-Chem.Phys.2001-p249,Stockburger-Phys.Rev.Lett.2002-p170407,Shao-J.Chem.Phys.2004-p5053}, which is a stochastic version of the Liouville-von Neumann equation without explicit environmental memory effect term in the equation. This equation can be applied to arbitrary spectral densities, but it may suffer from severe numerical instability in the long time simulation when the system-bath interaction is not small.
To overcome this problem, the idea of partial stochastic unraveling combined with partial hierarchical expansion was proposed, prompting the appearances of mixed deterministic-stochastic approaches\cite{Tanimura-J.Phys.Soc.Jpn.2006-p82001,Zhou-Europhys.Lett.2005-p334,Zhou-J.Chem.Phys.2008-p34106,
Moix-J.Chem.Phys.2013-p134106,Zhu-NewJ.Phys.2013-p95020}, which inherit both merits of the stochastic Liouville-von Neumann equation and hierarchical equations of motion.

Despite the great success, the aforementioned numerically exact methods are still limited in small sized systems due to the huge computational cost.
Towards the simulations of large-scale systems, one may turn to suitable approximate methods. Along with the history of quantum dynamics in open systems, quantum master equations play a prominent role\cite{Vega-Rev.Mod.Phys.2017-p}.
Starting from a formally exact quantum master equation\cite{Mohseni2014quantum,May-2008-p,breuer2002theory}, systematic approximations can be made to obtain computable equations of motion for different parameter regimes.
For example, in the weak exciton-phonon coupling regime, a perturbative treatment on the exciton-phonon coupling leads to the second-order time-convoluted or time-convolutionless quantum master equation, the Markovian limit of which is the famous Bloch-Redfield equation\cite{Redfield-1957theory,Redfield1965theory}.
Beyond the weak exciton-phonon coupling regime, the polaron transformation technique\cite{Jang2008-JCP-theory,Jang2009-JCP-theory,Nazir-Phys.Rev.Lett.2009-p146404,
Jang-J.Chem.Phys.2011-p34105,Kolli-J.Chem.Phys.2011-p154112,McCutcheon-Phys.Rev.B2011-p165101} and its variational version\cite{Silbey1984variational,Harris1985variational,Mccutcheon2011consistent,Mccutcheon2011general,
Lee2012accuracy,Pollock2013multi} are usually employed for a more accurate quantum master equation.

As compared with density matrix approaches, the non-Markovian stochastic Schr\"{o}dinger equation (NMSSE) has natural superiority in numerical calculations due to the beneficial linear scaling of Hilbert space.
About a few decades ago, Strunz, Di\'{o}si and coworkers\cite{Diosi-Phys.Lett.A1997-p569,Diosi-Phys.Rev.A1998-p1699,
Strunz-Phys.Rev.Lett.1999-p1801,Yu-Phys.Rev.A2004-p} pioneered the foundation of NMSSEs and proposed a formally exact NMSSE, the non-Markovian quantum state diffusion.
However, the explicit expression of the functional derivative term appearing in the equation is usually unknown, and one often invokes the version with zeroth order functional expansion approximation\cite{Yu-Phys.Rev.A1999-p91,Ritschel-NewJ.Phys.2011-p113034}.
Soon after the work of Strunz \etal, Gaspard and Nagoka\cite{Gaspard-J.Chem.Phys.1999-p5676,Vega-J.Chem.Phys.2005-p124106,Biele-J.Phys.:Condens.Matter2012-p273201} proposed a different NMSSE from the Feshbach projection method, which is only applicable to the situations when the system-bath interaction is weak and the bath relaxation is fast.
Lately another kind of NMSSE named the time-dependent wavepacket diffusion method was also proposed by Zhong and Zhao\cite{Zhong-J.Chem.Phys.2013-p14111}, and its high numerical performance was extensively demonstrated\cite{Zhong-J.Chem.Phys.2013-p14111,Zhong-NewJ.Phys.2014-p45009,
Han-J.Chem.Phys.2014-p214107,Ke-J.Phys.Chem.Lett.2015-p1741,Fujita-J.Phys.Chem.Lett.2016-p1374,Zang-J.Phys.Chem.C2016-p13351,
Plehn-J.Chem.Phys.2017-p34107,Zang2017quantum,Zhu2017charge}.
Very recently, inspired by the spirits of mixed deterministic-stochastic approaches, several studies showed that there exists a corresponding numerically exact hierarchical counterpart for a specific NMSSE.
Suess \etal\cite{Suess-Phys.Rev.Lett.2014-p150403,Suess-J.Stat.Phys.2015-p1408,
Ritschel-J.Chem.Phys.2015-p34115} applied the hierarchical expansion technique to the non-Markovian quantum state diffusion method and developed the hierarchy of pure states, which is proved to be highly efficient.
Song \etal\cite{Song-J.Chem.Phys.2016-p224105} utilized the same idea and obtained the hierarchical form of time-dependent wavepacket diffusion method.
Ke and Zhao developed a new hierarchical form of stochastic Schr\"{o}dinger equation\cite{Ke-J.Chem.Phys.2016-p24101}, and successively got its NMSSE version\cite{Ke-J.Chem.Phys.2017-p184103}. Various applications have shown the power of these NMSSEs and their hierarchical form in the simulations of complex systems dynamics.

In this paper, we are aimed at reviewing recent works about NMSSEs, their hierarchical forms and the applications.
In the next section, we start from a general partial stochastic unravelling scheme in the path-integral formalism, disentangling the forward and backward paths in the influence functional, which paves the way to the wavefunction-based framwork.
Then the hierarchical technique and systematic perturbation expansion are applied to the rest part of the influence functional, the former leads to the hierarchical form of stochastic Schr\"{o}dinger equations, whereas the latter leads to a set of NMSSEs.
In the third section, recent applications are presented to show the power of these approaches.
Concluding remarks are given in the last section.

\section*{\sffamily \Large Methodology}
\subsection*{\sffamily \large Background}
We start with a generic model where the total Hamiltonian consists of three parts:
\be
\label{tot-hamiltonian}
\hat{H}_{tot}=\hat{H}_{S}+\hat{H}_{B}+\hat{H}_{I}.
\ee
$\hat{H}_S$, $\hat{H}_B$, and $\hat{H}_I$ denote the system, the harmonic bath, and their interaction, respectively. Although the boundary between the system and its environment can be quite flexible, in molecular aggregates, like organic materials and photosynthetic systems, it is typical to choose the electronic degrees of freedom as the system part. The explicit form of $\hat{H}_B$ and $\hat{H}_I$ can be written as
\be
\hat{H}_{B}=\sum_{k}\frac{1}{2}(\hat{P}_{k}^{2}+\omega_{k}^{2}\hat{Q}_{k}^{2}),
\ee
and
\be
\hat{H}_{I}=\hat{x}\otimes\sum_{k}c_{k}\hat{Q}_{k},
\ee
where $\hat{P}_k$, $\hat{Q}_k$ and $\omega_k$ are the mass-weighted momentum, coordinate operator, and the frequency of the $k$-th bath mode, respectively, $\hat{x}$ is an operator characterizing the system-bath coupling manner and $c_k$ the coupling strength. In many cases, only the system dynamics is of interest, such that we would forward to obtain the reduced density operator by tracing over all the bath degrees of freedom. Assuming that the initial total density operator is factorized between the system and a thermal-equilibrium bath, i.e., $\hat{\rho}_{tot}(0)=\hat{\rho}_{S}(0)\otimes\hat{\rho}_{B}(0)$ with $\hat{\rho}_{B}(0)=e^{-\beta\hat{H}_{B}}/\trace\{e^{-\beta\hat{H}_{B}}\}$, where $\beta$ is the inverse temperature, the time-evolving reduced density operator of the system is given in the path-integral formalism as ($\hbar$ and $k_B$ are set to be unity for simplicity)
\be
\label{rhoS-orig}
\begin{split}
\rho_{S}(x_{f},x_{i},t)=&\la x_{f}|\trace_{B}\{e^{-i\hat{H}_{tot}t}\hat{\rho}_{tot}(0)e^{i\hat{H}_{tot}t}\}|x_{i}\ra  \\
=&\int \mathrm{d}x_{0}^{+}\int \mathrm{d}x_{0}^{-}\int_{x_{0}^{+}}^{x_{f}} \mathcal{D}x^+(t) \int_{x_{0}^{-}}^{x_{i}} \mathcal{D}x^-(t) e^{iS_0[x^+(t)]-iS_0[x^-(t)]}\rho_{S}(x_{0}^{+},x_{0}^{-},0)  \mathcal{F}[x^+(t),x^-(t),t],
\end{split}
\ee
where $\rho_{S}(x_{0}^{+},x_{0}^{-},0)=\la x_{0}^{+}|\hat{\rho}_{S}(0)|x_{0}^{-}\ra$, $S_0[x(t)]$ is the action functional corresponding to $\hat{H}_S$, and $\mathcal{F}[x^+(t),x^-(t),t]$ is the Feynman-Vernon influence functional\cite{Feynman-Ann.Phys.1963-p118} incorporating all the dynamical influences from the bath with the explicit form being
\be
\label{IF}
\mathcal{F}[x^+(t),x^-(t),t]=\exp\left\{-\int_{0}^{t}\mathrm{d}\tau\int_{0}^{\tau}\mathrm{d}\tau'\Phi(\tau,\tau')\right\},
\ee
and
\be
\label{phase}
\Phi(\tau,\tau')=[x^+(\tau)-x^-(\tau)][\alpha(\tau-\tau')x^+(\tau')-\alpha^{\ast}(\tau-\tau')x^-(\tau')].
\ee
Here $\alpha(t)$ is the bath correlation function, and its real and imaginary parts are denoted as $\alpha_{R}(t)$ and $\alpha_{I}(t)$, respectively. Defining the spectral density
\be
\label{SD}
J(\omega)=\frac{\pi}{2}\sum_{k}\frac{c_{k}^{2}}{\omega_{k}}\delta(\omega-\omega_{k}),
\ee
which totally characterizes the system-bath interaction, $\alpha(t)$ can be expressed as
\be
\label{alpha}
\begin{split}
\alpha(t)=&\alpha_{R}(t)+i\alpha_{I}(t) \\
=&\int_{0}^{\infty}\mathrm{d}\omega\frac{J(\omega)}{\pi}[\coth{\frac{\beta\omega}{2}}\cos{\omega t}-i\sin{\omega t}].
\end{split}
\ee

The complexity in the calculation of $\hat{\rho}_S(t)$ originates from the time-nonlocal nature of the influence functional as well as the entanglement between forward and backward paths in \Eq{IF}. Two powerful schemes, the stochastic unravelling\cite{Stockburger-Phys.Rev.Lett.2002-p170407,Shao-J.Chem.Phys.2004-p5053,Stockburger-Phys.Rev.Lett.1998-p2657,
Stockburger-J.Chem.Phys.1999-p4983,Stockburger-Chem.Phys.2001-p249} and deterministic hierarchical technique\cite{Tanimura-J.Phys.Soc.Jpn.1989-p101,Yan-Chem.Phys.Lett.2004-p216,Xu-J.Chem.Phys.2005-p41103,Ishizaki-J.Phys.Soc.Jpn.2005-p3131}, have been proposed to deal with these problems without any approximation. The total stochastic unravelling of the influence functional is straightforward, but its application to the realistic systems is limited by the convergence performance of stochastic average. Thus a variety of mixed deterministic-deterministic approaches have been proposed utilizing the partial stochastic unravelling strategy to improve the convergence speed\cite{Tanimura-J.Phys.Soc.Jpn.2006-p82001,Zhou-Europhys.Lett.2005-p334,Zhou-J.Chem.Phys.2008-p34106,
Moix-J.Chem.Phys.2013-p134106,Zhu-NewJ.Phys.2013-p95020}. In this review, we will follows this stream but with an additional requirement of the approaches capable of disentangling the forward and backward paths, leading to the wavefunction-based framework.

\subsection*{\sffamily \large Partial Stochastic Unravelling}
The partial stochastic unravelling requires an artificial partition of the influence functional, and the key to optimizing the convergence behavior of the stochastic ensemble average lies in this partition strategy. We first divide the bath correlation function $\alpha(t)$ into two parts
\be
\alpha(t)=\alpha^{1}(t)+\alpha_{res}(t),
\ee
with
\begin{subequations}
\label{alphaR-decom}
\be
\alpha^{1}(t)=\int_{0}^{\infty}\mathrm{d}\omega\frac{J(\omega)}{\pi}[2\bar{n}(\omega)+g(\omega)]\cos{\omega t},
\ee
\be
\alpha_{res}(t)=\int_{0}^{\infty}\mathrm{d}\omega\frac{J(\omega)}{\pi}\left\{\left[1-g(\omega)\right]\cos{\omega t}-i\sin\omega t\right\},
\ee
\end{subequations}
where $\bar{n}(\omega)=\frac{1}{e^{\beta\omega}-1}$ is the Bose distribution function, and $g(\omega)$ is a frequency-dependent function that will be specified later. Based on \Eq{alphaR-decom}, we further rewrite $\Phi(\tau,\tau')$ as the sum of a primary part and a residual part,
\be
\Phi(\tau,\tau')=\Phi_{p}(\tau,\tau')+\Phi_{r}(\tau,\tau'),
\ee
with
\be
\begin{split}
\Phi_{p}(\tau,\tau')=&x^+(\tau)\alpha_{R}^{1}(\tau-\tau')x^+(\tau')+x^-(\tau)\alpha_{R}^{1}(\tau-\tau')x^-(\tau')\\
&-x^+(\tau)\alpha^{\ast}(\tau-\tau')x^-(\tau')-x^-(\tau)\alpha(\tau-\tau')x^+(\tau')
\end{split}
\ee
and
\be
\Phi_{r}(\tau,\tau')=x^+(\tau)\alpha_{res}(\tau-\tau')x^+(\tau')+x^-(\tau)\alpha_{res}^\ast(\tau-\tau')x^-(\tau').
\ee

Note that the cross terms between $x^+(t)$ and $x^-(t)$ are entirely included in $\Phi_{p}(\tau,\tau')$. Thus if the partial stochastic unravelling is introduced for the corresponding part of the influence functional, the explicit entanglement between forward and backward paths can be resolved. One way towards the realization of stochastic unravelling is to use the Hubbard-Stratonovich transformation\cite{Hubbard-Phys.Rev.Lett.1959-p77,Stratonovich-1957-p416}. This results in an equivalent expression of \Eq{IF}:
\be
\label{IF-psu}
\mathcal{F}[x^+(t),x^-(t),t]=\int\mathcal{D}^{2}\chi_{1}\int\mathcal{D}^{2}\chi_{2}^{\ast}P[\chi_{1}(t),\chi_{2}^{\ast}(t),t]\exp^{-i\int_{0}^{t}[\chi_{1}(\tau)x^+(\tau)-\chi_{2}^{\ast}(\tau)x^-(\tau)]\mathrm{d}\tau-\int_{0}^{t}\mathrm{d}\tau\int_{0}^{\tau}\mathrm{d}\tau'\Phi_{r}(\tau,\tau')}.
\ee
Here, two correlated complex stochastic processes $\chi_1(t)$ and $\chi_2^\ast(t)$ are introduced, and $P[\chi_{1}(t),\chi_{2}^{\ast}(t),t]$ is the corresponding Gaussian probability density functional. By averaging over $\chi_1(t)$ and $\chi_2^\ast(t)$, \Eq{IF-psu} must recover to \Eq{IF}. To this end, it is found that $\chi_1(t)$ and $\chi_2^\ast(t)$ should satisfy the following statistical properties
\be
\label{sto-property}
\left\{
\begin{array}{l}
\average{\chi_{1}(t)}=\average{\chi_{2}^{\ast}(t)}=0,\\
\average{\chi_{1}(t)\chi_{1}(t')}=\average{\chi_{2}^{\ast}(t)\chi_{2}^{\ast}(t')}=\alpha_{R}^{1}(t-t'),\\
\average{\chi_{1}(t)\chi_{2}^{\ast}(t')}=\alpha^{\ast}(t-t').
\end{array}
\right.
\ee
In the coming subsections, we will provide the explicit generation scheme of $\chi_1(t)$ and $\chi_2^\ast(t)$ once the concrete form of $\alpha_R^1(t)$ is specified.

To go further, we need to specify the initial reduced density operator $\rho_S(0)$. Generally, one can always write it as $\hat{\rho}_{S}(0)=\sum_{i}|\psi_{1i}\ra\la \psi_{2i}|$. For simplicity, in the following we will assume that $\hat{\rho}_{S}(0)=|\psi_{1}\ra\la \psi_{2}|$, and the extension to the general cases just requires a simple summation. Substituting the initial condition and \Eq{IF-psu} into \Eq{rhoS-orig}, we can recast \Eq{rhoS-orig} into a compact form
\be
\label{rho-decom}
\rho_{S}(x_{f},x_{i},t)=\int\mathcal{D}^{2}\chi_{1}\int\mathcal{D}^{2}\chi_{2}^{\ast}P[\chi_{1}(t),\chi_{2}^{\ast}(t),t]\la x_f|\psi_{\chi_1}(t)\ra\la \psi_{\chi_2}(t)|x_i\ra.
\ee
The expressions of the forward stochastic wavefunction $|\psi_{\chi_1}(t)\ra$ and the backward stochastic wavefunction $| \psi_{\chi_2}(t)\ra$ are given by
\begin{subequations}
\label{AWV-0}
\label{psi-PI}
\be
\begin{split}
|\psi_{\chi_1}(t)\ra&=\int\mathcal{D}x(t)e^{iS_0[x(t)]-i\int_{0}^{t}x(\tau)
\left\{\chi_1(\tau)-i\int_{0}^{\tau}\alpha_{res}(\tau-\tau')x(\tau')\mathrm{d}\tau'\right\}\mathrm{d}\tau}|\psi_1\ra\\
&=\hat{\mathcal{U}}_{0}(t)\mathcal{T}_{X}^{\leftarrow}
\left[e^{-i\int_{0}^{t}\hat{X}(\tau)\left\{\chi_1(\tau)-i\int_{0}^{\tau}\alpha_{res}(\tau-\tau')\hat{X}(\tau')\mathrm{d}\tau'\right\}\mathrm{d}\tau}\right]|\psi_1\ra,
\end{split}
\ee
\be
\begin{split}
|\psi_{\chi_2}(t)\ra&=\int\mathcal{D}x(t)e^{iS_0[x(t)]-i\int_{0}^{t}x(\tau)
\left\{\chi_2(\tau)-i\int_{0}^{\tau}\alpha_{res}(\tau-\tau')x(\tau')\mathrm{d}\tau'\right\}\mathrm{d}\tau}|\psi_2\ra\\
&=\hat{\mathcal{U}}_{0}(t)\mathcal{T}_{X}^{\leftarrow}
\left[e^{-i\int_{0}^{t}\hat{X}(\tau)\left\{\chi_2(\tau)-i\int_{0}^{\tau}\alpha_{res}(\tau-\tau')\hat{X}(\tau')\mathrm{d}\tau'\right\}\mathrm{d}\tau}\right]|\psi_2\ra.
\end{split}
\ee
\end{subequations}
The second equality holds by realizing the fact that any product of multiple
time-dependent operators is automatically time-ordered in the path integral formalism\cite{Breuer-Phys.Rev.A2001-p,Ishizaki-Chem.Phys.2008-p185,
Ban-Phys.Lett.A2010-p2324}:
\be
\int_{x_i}^{x_f}\mathcal{D}x(t)e^{iS_0[x(t)]}x(t_n)\cdots x(t_1)\\
=\la X_f|\mathcal{T}_{X}^{\leftarrow}[\hat{X}(t_n)\cdots \hat{X}(t_1)]|X_i\ra, \quad t_f> t_n,\cdots,t_1>t_i
\ee
where $\hat{X}(t)=\hat{\mathcal{U}}_{0}^{\dagger}(t)\hat{x}\hat{\mathcal{U}}_{0}(t)$ and $|X_i\ra=\hat{\mathcal{U}}_{0}^{\dagger}(t_i)|x_i\ra$ with the definition of the pure system propagator $\hat{\mathcal{U}}_{0}(t)=e^{-i\hat{H}_{S}t}$. $\mathcal{T}_{X}^{\leftarrow}$ represents the chronological time-ordering operation of $\hat{X}(t_n)\cdots \hat{X}(t_1)$.

When an enough number of samplings of $|\psi_{\chi_1}(t)\ra$ and $|\psi_{\chi_2}(t)\ra$ have be accomplished, we can numerically evaluate $\hat{\rho}_S(t)$ via a statistical ensemble average according to \Eq{rho-decom}. But before that, by taking the time derivative of \Eq{psi-PI}, we obtain the following equation of motion in the operator form
\be
\label{EOM}
i\frac{\partial}{\partial t}|\psi_{\chi_j}(t)\ra=(\hat{H}_S+\chi_j(t)\hat{x})|\psi_{\chi_j}(t)\ra+\hat{x}\hat{\mathcal{U}}_{0}(t)\hat{\mathcal{R}}(\chi_j,t)
|\psi_j\ra\quad j=1,2.
\ee
with
\be
\label{AWV-1}
\hat{\mathcal{R}}(\chi_j,t)=\mathcal{T}_{X}^{\leftarrow}
\left[-i\int_{0}^{t}\alpha_{res}(t-\tau_1)\hat{X}(\tau_1)
\mathrm{d}\tau_1e^{-i\int_{0}^{t}\hat{X}(\tau_2)\left\{\chi_j(\tau_2)-i\int_{0}^{\tau_2}
\alpha_{res}(\tau_2-\tau_3)\hat{X}(\tau_3)\mathrm{d}\tau_3\right\}\mathrm{d}\tau_2}
\right].
\ee
Obviously, the hardcore issue of solving \Eq{EOM} resides in the treatment of the inhomogeneous term $\hat{x}\hat{\mathcal{U}}_{0}(t)\hat{\mathcal{R}}(\chi_j,t)
|\psi_j\ra$, which seems cumbersome at the first sight. In the following subsections, we will provide a few solutions using either the hierarchical technique or the perturbation expansion. The former leads to the numerically exact hierarchical form of stochastic Schr\"odinger equations, while the latter serves as the starting point to obtain a set of perturbative NMSSEs, which differ in the choice of $\alpha_{res}(t)$ and the level of approximations.

\subsection*{\sffamily \large Hierarchical Expansion}
When considering the cases where
\be
\label{alpha_assume}
\alpha_{res}(t)=\sum_{m=1}^M d_m e^{-\omega_m t},
\ee
one is able to construct a complete group consisting of an infinite number of auxiliary wavefunctions with the definition as
\be
\label{AWV-n}
|\Psi_{\chi_j}^{(\textbf{\emph{l}})}(t)\ra=\hat{\mathcal{U}}_{0}(t)\mathcal{T}_{X}^{\leftarrow}
\left[\prod_{m=1}^M\left(\int_{0}^{t}d_me^{-\omega_m(t-\tau_1)}\hat{X}(\tau_1)
\mathrm{d}\tau_1 \right)^{l_m}e^{-i\int_{0}^{t}\hat{X}(\tau_2)\left\{\chi_j(\tau_2)-i\int_{0}^{\tau_2}
\alpha_{res}(\tau_2-\tau_3)\hat{X}(\tau_3)\mathrm{d}\tau_3\right\}\mathrm{d}\tau_2}
\right]|\psi_j\ra.
\ee
which is closed with respect the time-derivative operation within this group. Besides, every frequency component is associated with an index number, leading to a vectorial index $\textbf{\emph{l}}=(l_1,l_2,\cdots,l_M)$ with $l_m$ being the non-negative integers.
Directly taking the time derivative of \Eq{AWV-n}, one obtains the following hierarchical set of differential equations of motion in a general form\cite{Suess-Phys.Rev.Lett.2014-p150403,Ke-J.Chem.Phys.2016-p24101,Song-J.Chem.Phys.2016-p224105}:
\be
\label{HSSE-general}
i\frac{\partial}{\partial t}|\Psi_{\chi_j}^{(\textbf{\emph{l}})}(t)\ra=\left(\hat{H}_S+\chi_j(t)\hat{x}-i\sum_{m=1}^{M}l_m\omega_m\right)
|\Psi_{\chi_j}^{(\textbf{\emph{l}})}(t)\ra+\sum_{m=1}^{M}l_md_m\hat{x}|\Psi_{\chi_j}^{(\textbf{\emph{l}}_m^-)}(t)\ra
+\hat{x}\sum_{m=1}^{M}|\Psi_{\chi_j}^{(\textbf{\emph{l}}_m^+)}(t)\ra,
\ee
where $\textbf{\emph{l}}_m^\pm=(l_1,\cdots,l_m\pm1,\cdots,l_M)$.
In the numerical simulations, it is truncated at a certain level $L=\sum_{m=1}^M l_m$. But obviously, the larger M is, the more wavefunctions are needed.
In the end, only the zeroth-tier term with $\textbf{\emph{l}}=(0,\cdots,0)$ are preserved for the stochastic ensemble average. By inspecting \Eq{AWV-n} and setting $t=0$, we obtain the initial condition for \Eq{HSSE-general}:
\be
|\Psi_{\chi_j}^{(\textbf{\emph{l}})}(0)\ra=
\left\{
\begin{array}{l}
|\psi_j\ra, \quad \textbf{\emph{l}}=(0,\cdots,0); \\
0, \quad else.
\end{array}
\right.
\ee

In the condensed phase, the bath consisting of nuclear vibrational modes and solvent degrees of freedom usually exhibits the overdamped features, thereby the spectral density can be represented by an Ohmic form with the Lorentzian cutoff\cite{The_Theory_of_Open_Quantum_Systems} (often called the Debye or Drude spectral density in the literatures)
\be
\label{SD-DB}
J(\omega)=\frac{2\lambda\omega\omega_c}{\omega^2+\omega_c^2},
\ee
where $\lambda$ and $\omega_c$ are the reorganization energy and the cutoff frequency, respectively. In this case, the corresponding bath correlation function can be analytically obtained as a series of exponential decaying functions with respect to time\cite{The_Theory_of_Open_Quantum_Systems}, satisfying the condition in \Eq{alpha_assume}.
\be
\label{alpha_debye}
\alpha(t)=\lambda\omega_{c}\coth{\frac{\beta\omega_c}{2}}e^{-\omega_ct}-\sum_{v=1}^{\infty}\frac{4\lambda\omega_c\omega_v}{\beta(\omega_c^2-\omega_v^2)}e^{-\omega_vt}
-i\lambda\omega_{c}e^{-\omega_{c}t},
\ee
where $\omega_v=2\pi v/\beta$ are the Matsubara frequencies. In the numerical implementations, a constraint on the upper bound of $\omega_v$ is indispensable in order to keep a finite index number. Nevertheless, it grows rapidly with the decreasing temperature.  Although a much more efficient frequency decomposition scheme, called Pad\'{e} spectrum decomposition has been proposed to alleviate this problem to some extent\cite{Hu-J.Chem.Phys.2010-p101106}, the numerical cost is still expensive, especially when the temperature is pretty low. Upon closer inspection of \Eq{alpha_debye}, it found that all the temperature effects are included in the real part. Therefore, we prefer to entirely circumvent this problem by choosing $g(\omega)=1$ in \Eq{alphaR-decom}, as a result, we have
\be
\alpha_{res}(t)=i\alpha_{I}(t)=-i\lambda\omega_{c}e^{-\omega_{c}t}.
\ee
In addition, an efficient generation scheme for $\chi_1$ and $\chi_2^\ast$ are introduced:
\begin{subequations}
\label{sto-HSSE-1}
\be
\chi_1(t)=\xi_c(t)+\xi_1(t),
\ee
\be\chi_2^\ast(t)=\xi_c^\ast(t)+\xi_2(t)
\ee
\end{subequations}
with the explicit expressions
\begin{subequations}
\label{sto-HSSE-2}
\be
\label{sto-HSSE-2-a}
\xi_c(t)=\sum_{k}\sqrt{\frac{c_k^2}{2\omega_k}}\left[\sqrt{\frac{n_k+1}{2}}(\mu_k^1+i\mu_k^2)e^{i\omega_kt} +\sqrt{\frac{n_k}{2}}(\mu_k^1-i\mu_k^2)e^{-i\omega_kt}\right],
\ee
\be
\xi_1(t)=\sum_{k}\sqrt{\frac{c_k^2}{2\omega_k}}(\sqrt{n_k+1}-\sqrt{n_k})
\times(\mu_k^3\cos{\omega_kt}+\mu_k^4\sin{\omega_kt}),
\ee
\be
\xi_2(t)=\sum_{k}\sqrt{\frac{c_k^2}{2\omega_k}}(\sqrt{n_k+1}-\sqrt{n_k})
\times(\mu_k^5\cos{\omega_kt}+\mu_k^6\sin{\omega_kt}),
\ee
\end{subequations}
where $n_k=\bar{n}(\omega_k)$, and $\mu_k^j$ ($j=1,\cdots,6$) are independent Gaussian white noises obey the normal distribution $N(0,1)$. Making use of the fact that $\average{\mu_k^i}=0$ and $\average{\mu_k^i\mu_{k'}^j}=\delta_{ij}\delta_{kk'}$, one can find that the generation scheme \Eq{sto-HSSE-1} and \Eq{sto-HSSE-2} indeed satisfies \Eq{sto-property} with $\alpha_{R}^{1}(t)$ being $\alpha_{R}(t)$.
Finally, \Eq{HSSE-general} is reduced to the following form:
\be
\label{HSSE}
i\frac{\partial}{\partial t}|\Psi_{\chi_j}^{(n)}(t)\ra=\left(\hat{H}_S+\chi_j(t)\hat{x}-in\omega_c\right)
|\Psi_{\chi_j}^{(n)}(t)\ra+\hat{x}|\Psi_{\chi_j}^{(n+1)}(t)\ra-in\lambda\omega_c\hat{x}|\Psi_{\chi_j}^{(n-1)}(t)\ra, \quad n=0,1,2,\cdots
\ee

\Eq{HSSE} possesses the merits of satisfactory statistical convergence performance, simplest hierarchical structure, and the favorable scaling property of Hilbert space, and so on. Thus, it is suitable to numerically exactly and efficiently explore the excitonic dynamics in Fenna-Matthews-Olson (FMO) trimer complexes of green sulfur bacteria\cite{Ke-J.Chem.Phys.2016-p24101} and peripheral light-harvesting complex 2 of the purple bacteria\cite{Ke-J.Chem.Phys.2017-p174105}, and other intermediate-sized systems. A recent study\cite{Ke2018quantum} showed that the method exhibits its strengths in simulating the system dynamics embedded in an ultraslow bath, where the non-Markovianity has proven to be extremely strong. In reality, there are a few numerical calculations\cite{Ke-J.Chem.Phys.2016-p24101,Ke2018quantum} demonstrating that the truncation level of hierarchical stochastic Schr\"odinger equation is generally less than that of the hierarchical equations of motion. One reason behind is that a majority of environmental influence has been taken into account through the introduction of stochastic fields, such that the higher-order phonon effects contained in the higher-order auxiliary wavefunctions are comparatively small. This has inspired us to overcome the limitations inherited in the hierarchy formalism, for example, the factorial scaling with respect to the system degrees of freedom and a few available spectral density functions, by resorting to the perturbative treatment\cite{Ke-J.Chem.Phys.2017-p184103}.

\subsection*{\sffamily \large Perturbation Expansion}

Let's refocus on Eqs.\,(\ref{EOM}) by reformulating it into a formally exact time-convolutionless equation of motion:
\be
\label{NMSSE-exact}
i\frac{\partial}{\partial t}|\psi_{\chi_j}(t)\ra=\left[\hat{H}_S+\chi_j(t)\hat{x}+\hat{x}\hat{\mathcal{U}}_0(t)\hat{\mathcal{K}}(\chi_j,t)\hat{\mathcal{U}}_{0}^{\dagger}\right]|\psi_{\chi_j}(t)\ra,
\ee
where the so-called residual dissipative generator $\hat{\mathcal{K}}(\chi_j,t)$ reads
\be
\label{kernel}
\begin{split}
\hat{\mathcal{K}}(\chi_j,t)=&\mathcal{T}_{X}^{\leftarrow}\left[-i\int_{0}^{t}\alpha_{res}(t-\tau_1)\hat{X}(\tau_1)
\mathrm{d}\tau_1e^{-i\int_{0}^{t}\hat{X}(\tau_2)\left\{\chi_j(\tau_2)-i\int_{0}^{\tau_2}\alpha_{res}(\tau_2-\tau_3)\hat{X}(\tau_3)\mathrm{d}\tau_3\right\}\mathrm{d}\tau_2}\right]\\
&\times\mathcal{T}_{X}^{\rightarrow}\left[e^{i\int_{0}^{t}\hat{X}(\tau_1)\left\{\chi_j(\tau_1)-i\int_{0}^{\tau_1}\alpha_{res}(\tau_1-\tau_2)\hat{X}(\tau_2)\mathrm{d}\tau_2\right\}\mathrm{d}\tau_1}\right],
\end{split}
\ee
where $\mathcal{T}_{X}^{\rightarrow}$ denotes the anti-chronological time-ordering operation. This expression is the starting point of the systematic perturbation expansion with respect to the system-bath coupling strength $c_k$. It can be seen from \Eq{SD}, \Eq{alpha} and \Eq{sto-property} that $\chi_j(t)$ is of the first order of $c_k$, while $\alpha(t)$ is of the second order. By expanding the two exponential functions in \Eq{kernel} directly and paying particular attention to the time-ordering operation, we can obtain the perturbation expansion of \Eq{kernel} up to arbitrary order as $\hat{\mathcal{K}}(\chi_j,t)=\sum_{n=2}^{\infty}\hat{\mathcal{K}}_{n}(\chi_j,t)$. The lowest (second) order therein is  explicitly expressed as
\be
\label{kernel-2}
\hat{\mathcal{K}}_2(t)=-i\int_{0}^{t}\alpha_{res}(t-\tau)\hat{X}(\tau)\mathrm{d}\tau.
\ee
As a matter of fact, It is advisable to approximate the residual dissipative generator as $\hat{\mathcal{K}}(\chi_j,t)\approx\hat{\mathcal{K}}_2(t)$, as we have affirmed in a numerically manner that the contributions of higher order terms are generally small over a quite broad parameter space\cite{Ke-J.Chem.Phys.2017-p184103}. Inserting \Eq{kernel-2} into \Eq{NMSSE-exact} and then making the substitution $t-\tau\rightarrow\tau$, we obtain the final result of this subsection:
\be
\label{NMSSE-general}
i\frac{\partial}{\partial t}|\psi_{\chi_j}(t)\ra=\left[\hat{H}_S+\chi_j(t)\hat{x}-i\hat{x}\int_{0}^{t}\alpha_{res}(\tau)e^{-i\hat{H}_{S}\tau}\hat{x}e^{i\hat{H}_{S}\tau}\mathrm{d}\tau\right]|\psi_{\chi_j}(t)\ra.
\ee

A significant point about \Eq{NMSSE-general} is that it is applicable to arbitrary form of spectral densities, which is very important since more and more studies have stressed that the realistic highly structured environment might play a crucial role in various cases like the enhancement of energy transfer rate\cite{Rey2013exploiting,OReilly-Nat.Commun.2014-p,Dijkstra2015coherent,Iles2016energy} and non-Markovianity\cite{Rebentrost-J.Chem.Phys.2011-p101103,Mujica-Martinez-Phys.Rev.E2013-p,
Vaughan_J.Chem.Phys._2017_p124113}, the long-sustained quantum coherence in light-harvesting complexes\cite{Christensson2012origin,Tiwari2013electronic,Chenu2013enhancement,Chin-Nat.Phys.2013-p113}, and the fine-tuning of the dephasing and relaxation time in quantum information processors\cite{Wilhelm-Chem.Phys.2004-p345,Thorwart-Chem.Phys.2004-p333}. What's more, since the concrete form of $\alpha_{res}(t)$ which exerts a substantial influence on the numerical performance of \Eq{NMSSE-general}, has not yet been specified, this flexibility allows us to seek for a balance point between high accuracy and affordable computational cost.

After tons of numerical simulations based on the spin-boson model over a broad parameter range, it is found that an excellent numerical performance is guaranteed by the following choice of the pairwise $\chi_j(t)$ and $\alpha_{res}(t)$:
\begin{subequations}
\label{tdwpd}
\be
\chi_1(t)=\chi_2(t)=\xi_c(t), \quad \text{as shown in \Eq{sto-HSSE-2-a} };
\ee
\be
\alpha_{res}(t)=\int_{0}^{\infty}\mathrm{d}\omega\frac{J(\omega)}{\pi}[\tanh{\frac{\beta\omega}{4}}\cos{\omega t}-i\sin{\omega t}].
\ee
\end{subequations}
Specifically, we have denoted \Eq{NMSSE-general} together with the definitions of $\chi_{1/2}(t)$ and $\alpha_{res}(t)$ shown in \Eq{tdwpd} as the perturbative stochastic Schr\"odinger equation (PSSE) in the coming numerical section. The numerical efficiency of the PSSE method has been displayed by comparing with that of other versions of SSEs, benchmarked against the exact results from \Eq{HSSE-general} for a wide range of parameters in the spin-boson model\cite{Ke-J.Chem.Phys.2017-p184103}, and it was found to be relatively stable from weak to intermediate system-bath coupling regimes at different temperatures and bath characteristic frequencies. We will further affirm its valid regime in a simple dimer model before it is applied to realistic organic aggregates in this paper. In the past few years, the PSSE method has proven its strength in simulating charge carrier dynamics in large-scale realistic organic photovoltaic systems composed of a few hundreds (thousands) monomers\cite{Zhong-NewJ.Phys.2014-p45009,
Han-J.Chem.Phys.2014-p214107,Ke-J.Phys.Chem.Lett.2015-p1741,Fujita-J.Phys.Chem.Lett.2016-p1374,Zang-J.Phys.Chem.C2016-p13351,
Plehn-J.Chem.Phys.2017-p34107,Zang2017quantum,Zhu2017charge}. Some specific examples will be detailedly presented in the next section.

In some limiting cases, for instance, highly-symmetric systems and extremely high-temperature conditions, it is reasonable to further invoke a crude approximation\cite{Zhong-J.Chem.Phys.2011-p134110}, i.e.,
\be
\label{crude-approx}
\alpha(t)\approx\alpha_{R}(t),
\ee
which actually implies a semiclassical assumption. Under this circumstance, \Eq{NMSSE-exact} is greatly simplified  to
\be
\label{oTDWPD}
i\frac{\partial}{\partial t}|\psi_{\chi_j}(t)\ra=\left(\hat{H}_{S}+\widetilde{\xi}_c(t)\hat{x}\right)|\psi_{\chi_j}(t)\ra,
\ee
since $\alpha_{res}(t)=0$ and the stochastic process $\widetilde{\xi}_c(t)$ can be reduced to the real one
\be
\label{sto-oTDWPD}
\widetilde{\xi}_c(t)=\sum_{k}\sqrt{\frac{c_k^2}{\omega_k}\coth{\frac{\beta\omega_k}{2}}}\left(\mu_k^1\cos(\omega_kt)+\mu_k^2\sin(\omega_k t)\right),
\ee
where $\mu_k^1$ and $\mu_k^2$ are independent real Gaussian random variables obeying the standard normal distribution, and it is trivial to verify that $\average{\widetilde{\xi}_c(t)}=0$ and $\average{\widetilde{\xi}_c(t)\widetilde{\xi}_c(t')}=\alpha_{R}(t-t')$. If the bath response time is ultrashort in comparison with the system dynamics,  one can proceed to approximate $\alpha_R(t)$ as a delta function (white noise), such that \Eq{oTDWPD} recovers the classical Haken-Strobl-Reineker model (HSR)\cite{Haken1972coupled,Haken1973exactly,Reineker1981drift}, and out of this reason, \Eq{oTDWPD} is referred to as the modified-HSR method in the following numerical simulations. Even though the neglect of $\alpha_I(t)$ leads to the break down of detailed balance condition, as long as one can take care of its range of application in the practical simulations,  \Eq{oTDWPD} still constitutes a valuable and reliable simulation tool widely-used in nanoscale systems. For instance, it is utilized in evaluating and predicting the charge carrier mobilities in many representative organic materials\cite{Zhang2012electron,Jiang2016nuclear}, and is used to investigate the effects of spatial as well as temporal correlations between site energies on the charge transfer (CT) process in flexible molecules\cite{Liu2016coarse}. More interestingly, \Eq{oTDWPD} toghether with the help of massively parallel computing platforms\cite{Sawaya-NanoLett.2015-p1722} and  advanced experimental techniques, is capable to set up a bottom-up strategy\cite{Boulais2018programmed} promising for the optimal design of advanced organic functional materials.

\section*{\sffamily \Large Applications}
In this section, we will present a few representative applications of the aforementioned methods.
Various numerical algorithms like the forth-order Runge-Kutta method and Chebyshev polynomial expansion
are applicable to the propagation of stochastic Schr\"{o}dinger equations. It should be noted that the norm of the stochastic wavefunction is actually not conserved during the propagation, so an artificial renormalization is generally performed in the practical simulations. However, it is possible to transform the linear stochastic Schr\"{o}dinger equation into a norm-conserved nonlinear stochastic Schr\"{o}dinger equation by means of Girsanov transformation. Various nonlinear forms of \Eq{HSSE-general} and \Eq{NMSSE-general} can be found in the literatures\cite{Suess-Phys.Rev.Lett.2014-p150403,Ke-J.Chem.Phys.2016-p24101,Diosi-Phys.Rev.A1998-p1699,
Yu-Phys.Rev.A1999-p91,Plehn-J.Chem.Phys.2017-p34107}. In the hierarchy of pure state method (HOPS)\cite{Suess-Phys.Rev.Lett.2014-p150403}, the non-linear version is reported to remarkably accelerate the convergence speed, while this conclusion does not apply to \Eq{HSSE}\cite{Ke-J.Chem.Phys.2016-p24101}. Concerning the fact that linear SSE is computationally beneficial than its non-linear correspondence, especially for large systems, we are prone to using the linear form in the following simulations.

Up to now, the hierarchical form of stochastic Schr\"odinger equations have been used in simulating the exciation energy transfer in full 24-site FMO complexes\cite{Ke-J.Chem.Phys.2016-p24101}, the linear spectra of light-harvesting complex 2\cite{Ke-J.Chem.Phys.2017-p174105}, as well as investigating the non-Markovianity of a system embedded in an ultraslow bath\cite{Ke2018quantum}. In the following, we will demonstrate in great details the superiority of the HSSE method over its deterministic counterpart, the HEOM method with regards to the simulations of excitation energy transfer in large-scale photosynthetic systems.

\subsubsection*{\sffamily \normalsize Energy Transfer in the FMO Complex}
Fenna-Matthews-Olson complex is a light harvesting antennae found in green sulfur bacteria, bridging a large chromosome supercomplex to the reaction center. Due to its relatively small size and well-resolved network structure, it has long served as a prototypical model to explore the highly efficient energy transfer in photosynthesis and has attracted tremendous attention from numerous research areas, including various experimental and theoretical spectroscopic studies \cite{Brixner-Nature2005-p625,Engel-Nature2007-p782,
Panitchayangkoon-Proc.Natl.Acad.Sci.USA2010-p12766,Cho-J.Phys.Chem.B2005-p10542,Chen-J.Chem.Phys.2011-p194508,Hein-NewJ.Phys.2012-p23018,
Kreisbeck-J.Phys.Chem.Lett.2012-p2828,duan2017nature}, atomistic simulations\cite{Olbrich-J.Phys.Chem.B2011-p8609,Adolphs-Biophys.J.2006-p2778,Gao-J.Phys.Chem.B2013-p3488}, electronic structure and quantum dynamics calculations\cite{Ishizaki-Proc.Natl.Acad.Sci.U.S.A.2009-p17255,Moix-J.Phys.Chem.Lett.2011-p3045,
Kreisbeck-J.Chem.TheoryComput.2011-p2166,Mujica-Martinez-Phys.Rev.E2013-p,
Nalbach_PRE_15_022706,Schulze-J.Chem.Phys.2016-p185101,Ritschel-NewJ.Phys.2011-p113034}, and so on. Two-dimensional electronic spectroscopy studies of FMO complex during the past decade have brought an unprecedentedly high time- and frequency-resolution of the underlying ultrafast physical processes, while the consensus about the interpretation of observed unexpectedly long-lived peak oscillating behaviors has not yet been reached\cite{Engel-Nature2007-p782,Panitchayangkoon-Proc.Natl.Acad.Sci.USA2010-p12766,Christensson2012origin,Tiwari2013electronic,
Kreisbeck-J.Phys.Chem.Lett.2012-p2828,Chenu2013enhancement,Ishizaki-Proc.Natl.Acad.Sci.U.S.A.2009-p17255,Chin-Nat.Phys.2013-p113,
duan2017nature,Mohseni2014quantum,Lambert-Nat.Phys.2013-p10,Scholes-Nat.Chem.2011-p763}. Thus, in the quantum dynamics branch, it would be preferable to use the methods as accurate as possible. The first non-perturbative  quantum dynamics calculation of the excitation energy transfer in FMO complex was performed by Ishizaki and Fleming\cite{Ishizaki-Proc.Natl.Acad.Sci.U.S.A.2009-p17255} using the HEOM approach, based on a well-established excitonic Hamiltonian for a seven-bacteriochlorophyll subunit out of three-fold circularly symmetric assembly. It is further assumed that every pigment is coupled to its own environment which is characterized by the Debye spectral density function, as shown in \Eq{SD-DB}. Based on the same model, we would exemplify the applicability and efficiency of the HSSE method, in comparison with the popular HEOM method.
\begin{figure}
  \centering
  \includegraphics[width=0.7\textwidth]{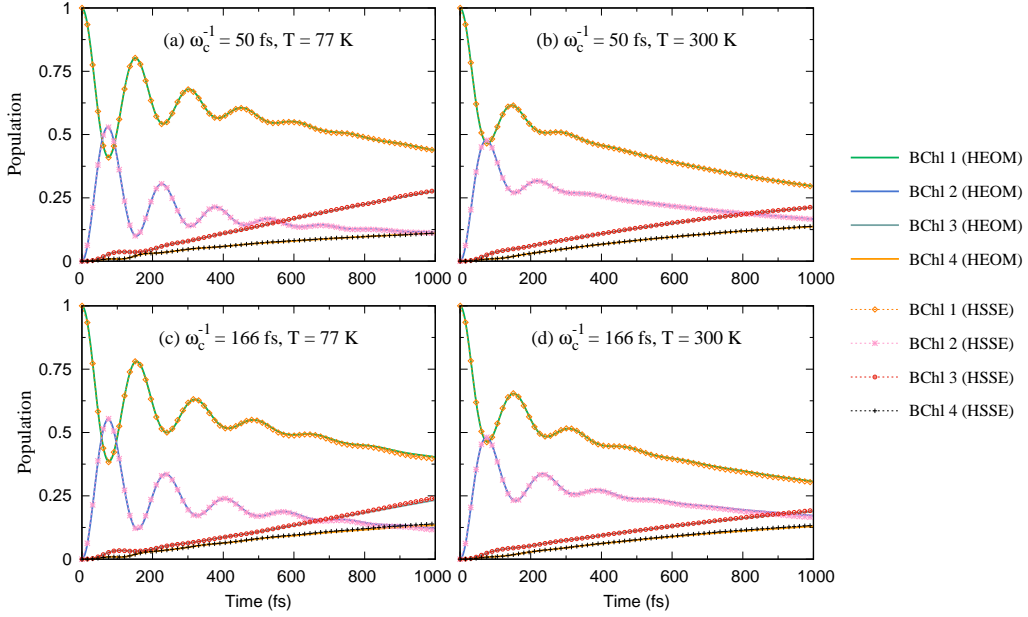}\\
  \caption{Plots of time-evolving populations at BChl 1, 2, 3, and 4 for a seven-site subunit of FMO complex, under four different parameter sets with the specific values given at the top of every panel. The reorganization energy is fixed at $\lambda=35\,\mathrm{cm}^{-1}$. The initial population is located at BChl 1.}\label{FMO-7site}
\end{figure}
\begin{table}
  \centering
  \begin{tabular}{c||c|c||c|c}
     \hline\hline
     Parameter Set &\multicolumn{2}{c||}{\textbf{HEOM}} & \multicolumn{2}{c}{\textbf{HSSE}} \\ \cline{2-5}
      $(\omega_c^{-1},\, \mathrm{T})$ & $(\mathrm{L_{HEOM}},\,\mathrm{K})$ & (\#AM)$\times \mathrm{N}^2$ & $(\mathrm{L_{HSSE}},\,\mathrm{N_{ran}})$ & (\#AW)$\times$ 2N \\ \hline
     $(50\mathrm{fs},\, 77\mathrm{K})$ & $(4,\,3)$ &    619850 & $(3,\,20000)$ & 1680 \\ \hline
     $(50\mathrm{fs},\, 300\mathrm{K})$ & $(5,\,1)$ &    38808 & $(3,\,10000)$ & 1680\\ \hline
     $(166\mathrm{fs},\, 77\mathrm{K})$ & $(6,\,2)$ &  1899240 & $(3,\,10000)$ & 1680 \\ \hline
     $(166\mathrm{fs},\, 300\mathrm{K})$ & $(10,\,1)$ & 952950 & $(3,\,5000)$ & 1680 \\
     \hline
   \end{tabular}
  \caption{The computational requirements for the HEOM and HSSE methods to obtain the converged results as shown \Fig{FMO-7site}. $\mathrm{L_{HEOM}}$ and $\mathrm{L_{HSSE}}$ are the truncation levels of hierarchy for HEOM and HSSE method, respectively. $\mathrm{K}$ is the number of Matsubara frequencies needed and \#AM denotes the number of auxiliary matrices in HEOM method. $\mathrm{N_{ran}}$ is the random trajectory number to obtain smooth curves and \#AW denotes the number of auxiliary wavefunctions used in HSSE method. Besides, $\mathrm{N}=7$ in this case.}\label{tab1}
\end{table}
\begin{table}
  \centering
  \begin{tabular}{c||c|c||c|c}
     \hline\hline
     Parameter Set &\multicolumn{2}{c||}{\textbf{HEOM}} & \multicolumn{2}{c}{\textbf{HSSE}} \\ \cline{2-5}
      $(\omega_c^{-1},\, \mathrm{T})$ & $(\mathrm{L_{HEOM}},\,\mathrm{K})$ & (\#AM)$\times \mathrm{N}^2$ & $(\mathrm{L_{HSSE}},\,\mathrm{N_{ran}})$ & (\#AW)$\times$ 2N \\ \hline
     $(50\mathrm{fs},\, 77\mathrm{K})$ & $(4,\,3)$ & 738993600 & $(3,\,20000)$ & 140400 \\ \hline
     $(50\mathrm{fs},\, 300\mathrm{K})$ & $(5,\,1)$ & 68402880 & $(3,\,10000)$ & 140400\\ \hline
     $(166\mathrm{fs},\, 77\mathrm{K})$ & $(6,\,2)$ &   14876447040 & $(3,\,10000)$ & 140400 \\ \hline
     $(166\mathrm{fs},\, 300\mathrm{K})$ & $(10,\,1)$ & 75529808640 & $(3,\,5000)$ & 140400 \\
     \hline
   \end{tabular}
\caption{The computational requirements for the HEOM and HSSE methods to obtain the converged results for full 24-site FMO trimer. The meaning of the characters are the same as those in Table. \ref{tab1} except that $\mathrm{N}=24$ in this case.}\label{tab2}
\end{table}
\begin{figure}
  \centering
  \includegraphics[width=0.7\textwidth]{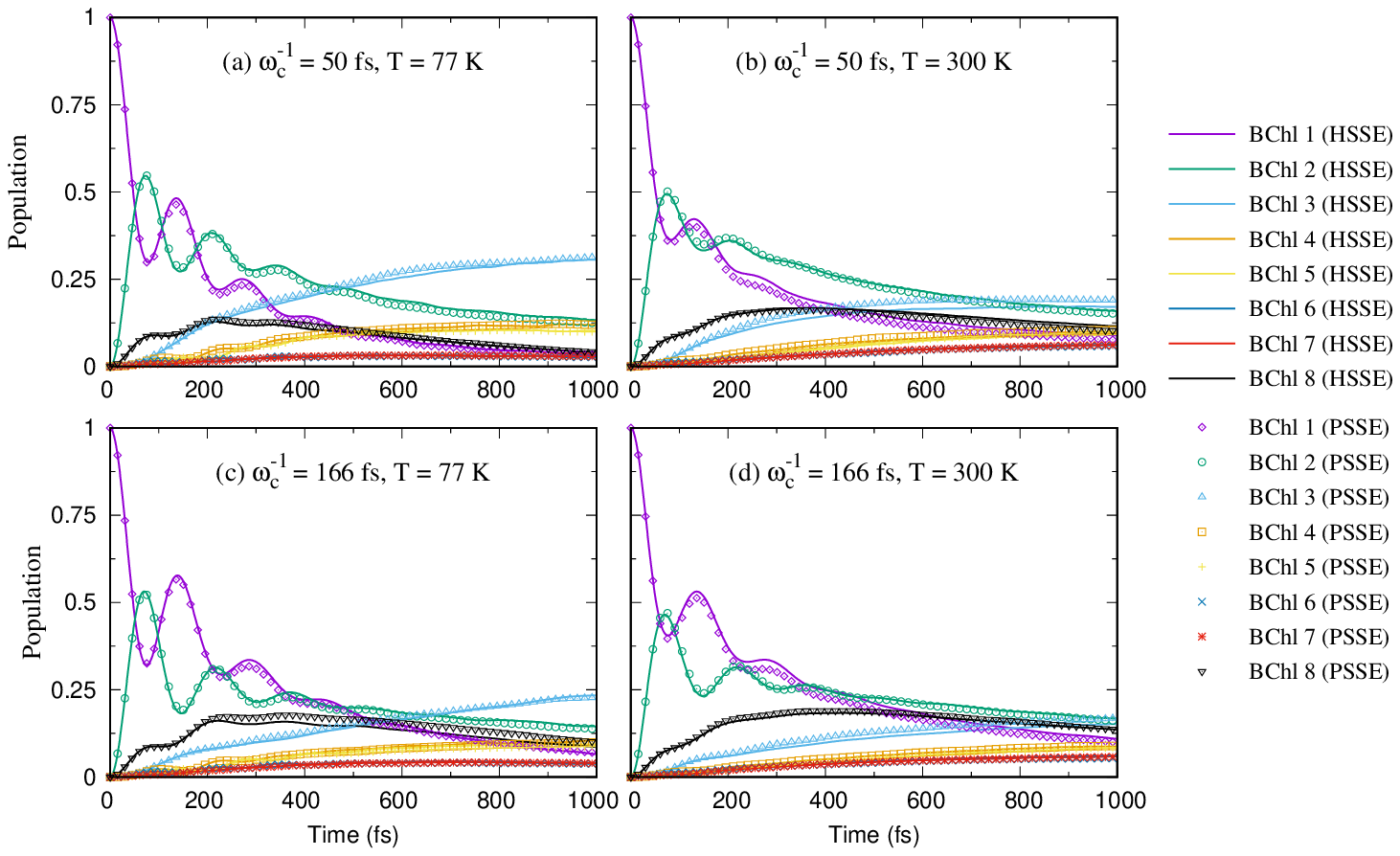}\\
  \caption{Plots of time-evolving populations for a full 24-site FMO complex, under four different parameter sets with the specific values given at the top of every panel. The reorganization energy is fixed at $\lambda=35\,\mathrm{cm}^{-1}$. The initial population is located at BChl 1 within one subunit out of a homotrimer. For the purpose of a better visualization, only those corresponding to the most populated eight BChls are displayed.}\label{FMO-24site}
\end{figure}

Shown in \Fig{FMO-7site} are the population evolutions of the first four bacteriochlorophylls (BChl) obtained through both the HEOM and HSSE methods, under four different parameter sets of the temperature $\beta$ and bath cutoff frequency $\omega_c$, while the bath reorganization energy is fixed at $\lambda=35\,\mathrm{cm}^{-1}$. The initial excitation is prepared at BChl 1. The numerical conditions for convergence are listed in Table. \ref{tab1}. Note that the higher the temperature or the smaller $\omega_c$ is, the deeper the truncation level $\mathrm{L_{HEOM}}$ is needed in HEOM method. $\mathrm{K}$ will be larger as well at the lower temperature. In contrast, $\mathrm{L_{HSSE}}$ seems less susceptible to the temperature and bath cutoff frequency\cite{Ke2018quantum}. Besides, as mentioned in previous section about HSSE (\Eq{HSSE}), all the temperature effects are included in the stochastic noises and consequently neither Matsubara nor Pad\'e frequency decomposition schemes are needed. In the end, the number of elements needed to be propagated in HSSE is far smaller than that in HEOM method, which indicates a substantial saving of memory requirements. Typically in photosynthetic systems or organic aggregates, the electron-phonon interaction is comparatively small or in close proximity to the excitonic couplings, such that the random trajectory number $\mathrm{N_{ran}}$ in most cases remains manageable. In passing, the stochastic nature renders HSSE method well-suited for parallel computing and allows for the inclusion of static disorders in a trivial fashion\cite{Fokas-J.Phys.Chem.Lett.2017-p2350}. The advantages will be more evident in larger complexes.

Very recently, it is found that there exits in vivo an additional BChl, termed as BChl 8, residing in a cleft at the surface of the protein to connect two neighboring subunits\cite{Tronrud-Photosynth.Res.2009-p79,SchmidtamBusch-J.Chem.TheoryComput.2010-p93,Olbrich-J.Phys.Chem.B2011-p8609}. As such, one has to consider the intact trimer structure composed of 24 BChls. The computational requirements grow a lot for both HEOM and HSSE methods, but obviously more demanding for the former, as illustrated in Table. \ref{tab2}. The results of population evolutions of full trimer at the bath condition $\omega_c^{-1}=50\,\mathrm{fs}$ and $\mathrm{T}=300\,\mathrm{K}$ are calculated by Wilkins and Dattani\cite{Wilkins-J.Chem.TheoryComput.2015-p3411} through the HEOM method with the usage of a more efficient numerical propagation algorithm. Nevertheless, applications to the rest three parameter sets are far more challenging. For the most straightforward example, at the case $\omega_c^{-1}=166\,\mathrm{fs}$ and $\mathrm{T}=300\,\mathrm{K}$, up to 75 billions elements are needed to be propagated in a single time step. While, complementarily, the numerical expenses for the HSSE method in this parameter regime appear to be the cheapest. The converged results acquired by the HSSE method under four different sets of bath parameters are shown in \Fig{FMO-24site}, with the excitonic Hamiltonian give by
\be
\label{He-FMO}
\hat{H}_{ex}=
\left[
\begin{array}{ccc}
  \hat{H}^1_A    &     \hat{H}_B     &   \hat{H}_B^T   \\
  \hat{H}_B^T  &     \hat{H}^1_A     &   \hat{H}_B     \\
  \hat{H}_B    &     \hat{H}_B^T   &   \hat{H}^1_A     \\
\end{array}
\right],
\ee
$H_A^1$ covers the intra-subunit excitonic elements\cite{Jia-Sci.Rep.2015-p17096} and $H_B$ represents all the inter-subunit couplings\cite{Olbrich-J.Phys.Chem.B2011-p8609}, and they are specifically tabulated in Ref. \cite{Ke-J.Chem.Phys.2016-p24101}.
In addition, we have also plotted the results obtained through the PSSE method, i.e., \Eq{NMSSE-general}, in \Fig{FMO-24site}. Benchmarked by the exact results, the PSSE method has proven to be remarkably accurate for delineating the population relaxation in FMO complexes. In fact, the simulations using modified-HSR method (\Eq{oTDWPD}) are also performed, but the results are qualitatively wrong and then not shown here. Since \Eq{NMSSE-general} and \Eq{oTDWPD} are applicable to nanoscale realistic systems consisting of several thousands and tens of thousands of system degrees of freedom, respectively. It is quite essential to systematically investigate their ranges of validity beforehand.

\subsubsection*{\sffamily \normalsize Validity Regime of Two Approximate SSEs}
\begin{figure}
  \centering
  \includegraphics[width=0.85\textwidth]{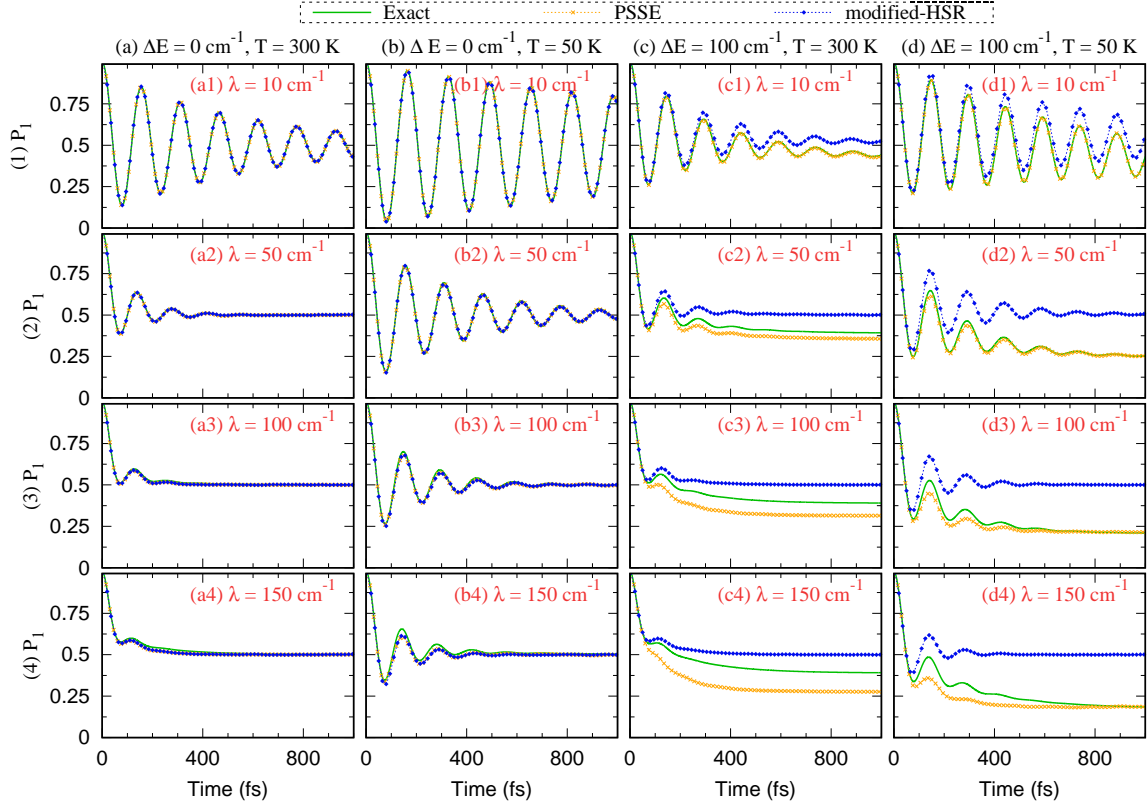}\\
  \caption{Plots of the time-evolving populations at site 1 in a dimer model system, subjected to various reorganization energies $\lambda$ $\left[(1)-(4)\right]$ and two different energy biases $\big[\Delta E=E_1-E_2=0$ and $\Delta E=100\,\mathrm{cm}^{-1}\big]$ at both high and low temperatures $\big[\mathrm{T}=300\,\mathrm{K}$ and $\mathrm{T}=50\,\mathrm{K}\big]$. The initial population is put at site 1. Other parameters are set as: the excitonic coupling $V=100\,\mathrm{cm}^{-1}$, $\omega_c^{-1}=100\,\mathrm{fs}$ in the Debye spectral density function \Eq{SD-DB}.}
  \label{Dimer}
\end{figure}
The perturbative forms of stochastic Schr\"odinger equations, including \Eq{NMSSE-general} and \Eq{oTDWPD}, due to their favorable linear scaling to the system size, well-behaved stochastic convergence property, and the time-convolutionless structure, have proven to be extremely suitable for describing the quantum dynamics of various large-scale organic systems and have achieved remarkable success, for example, successfully illuminating the temperature dependence of carrier diffusion coefficients in one-dimensional chain models including both static and dynamic disorders for a wide range of disorder strength, and bridging the band-like and hopping-type transport\cite{Zhong-NewJ.Phys.2014-p45009}, demonstrating the effects of the laser pulse, the Coulomb interaction and the temperature in the energy relaxation as well as the separation of a photon-induced hot electron-hole pair in organic aggregates\cite{Han-J.Chem.Phys.2014-p214107}. Besides, it can be straightforwardly extended to investigate  the mixed Frenkel exciton and CT exciton states, and then demonstrated the important role of hole delocalization on the Frenkel-CT decoherence time\cite{Fujita-J.Phys.Chem.Lett.2016-p1374}.

It is recommendable to testify the validity of approaches on the basis of a simple model subjected to various parameter conditions. The simplest molecular aggregate, namely, the dimer, is adopted here, while some key features should be shared by larger analogous systems. As was found previously in Ref. \cite{Ke-J.Chem.Phys.2017-p184103} and also demonstrated in the FMO complex shown above, the accuracy of the PSSE and modified-HSR methods seems barely influenced by the cutoff frequency. As such, we have fixed it at the value $\omega_c^{-1}=100\,\mathrm{fs}$ in the calculations. The excitonic coupling is chosen as a reference and set as $V=100\,\mathrm{cm}^{-1}$.
In a homogeneous sample with the energy biases being zeros, as shown in \Fig{Dimer} (a) and (b), the results of both PSSE and modified-HSR methods are in good agreement with the exact ones, even when the reorganization energy $\lambda$ is comparable to or slightly larger than the excitonic couplings. In fact, the results remain qualitatively solid till $\lambda=3\,V$. Nevertheless, this range is greatly reduced when the finite energy bias $(\Delta E=E_1-E_2\neq 0)$ is introduced. It could be concluded that the modified-HSR method is completely not suited for an inhomogeneous system. As can clearly seen in \Fig{Dimer} (c) and (d), it fails even under very weak system-bath coupling condition $(\lambda/V=0.1)$, and the reason behind is that the imaginary part of bath correlation function is completely neglected, that is, all the feedbacks from the environment to the system are missed and thus the detailed balance is broken down. The PSSE method has partially repaired this disability, especially at the low temperatures. In short, the PSSE method might overestimate the relaxation rates when applied to an inhomogenous system, but generally speaking, the results should be trustworthy when $\lambda/V < 1$.

The rest two subsections mainly introduce the applications of the PSSE method (\Eq{NMSSE-general}) to unveil the multiple time scales hot exciton relaxation process in a neat organic aggregate\cite{Ke-J.Phys.Chem.Lett.2015-p1741}, and study the CT states, exciton migrations, and the quantum interference effects in singlet fission (SF) dynamics\cite{Zang2017quantum}.

\subsubsection*{\sffamily \normalsize Hot Exciton Relaxation}
The energy relaxation of a hot exciton (an excitonic state with energy exceeding $k_BT$ considerably) is believed to be critical to many dynamic processes in organic photovoltaics like charge separation and multiple exciton generation. Recent experiments found that the excess energy of the hot exciton goes through an initial fast followed by a slow relaxation process\cite{Yang-Phys.Rev.B2005-p45203,Engel-Nature2007-p782,Banerji-J.Phys.Chem.C2011-p9726,
Scheblykin-J.Phys.Chem.B2007-p6303,Kee-J.Phys.Chem.Lett.2014-p3231,Dai-Phys.Rev.B2013-p45308,
Wells-Phys.Rev.Lett.2008-p86403,Kempe-Contemp.Phys.2003-p307}. Further studies indicate that the fast one, which may be relevant to the coherent motion of the exciton, can largely enhance the generation of separated charges at the donor-acceptor interface\cite{Kaake-J.Phys.Chem.Lett.2013-p2264}. However, the detailed mechanism of the hot exciton energy relaxation is still not clear. A recent work\cite{Ke-J.Phys.Chem.Lett.2015-p1741} have given an unambiguous picture of this process and clarified the ultimate origin of the multiple time scales processes.

\begin{figure}[H]
	\centering
	\includegraphics[width=0.5\textwidth]{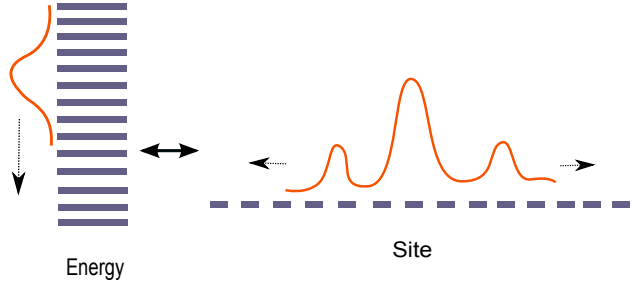}
	\caption{Transform of initial distribution from the energy representation to the site representation. The energy band ranges from $-2V_{i,i\pm 1}$ to $2V_{i,i\pm 1}$ as
  the excitonic energies $\{\epsilon_n\}$ are all set as zero. Reprinted with permission from \rcite{Ke-J.Phys.Chem.Lett.2015-p1741}. Copyright 2015 American Chemical Society.}
	\label{fig1-he}
\end{figure}

Here we consider the Frenkel exciton model for an one-dimensional aggregate chain. The Hamiltonian of the exciton part is
\be
\hat{H}_{ex}=\sum_{n}\epsilon_{n}\hat{B}_{n}^{\dagger}\hat{B}_{n}+\sum_{n>m}V_{nm}(\hat{B}_{n}^{\dagger}\hat{B}_{m}+\hat{B}_{m}^{\dagger}\hat{B}_{n}),
\ee
where $\hat{B}_{n}^{\dagger}$ and $\hat{B}_{n}$ are the creation and annihilation operator of a local exciton at the nth site, and $\epsilon_n$ is the corresponding energy. $V_{nm}$ is the excitonic coupling between mth and nth sites, which can be calculated through the Coulomb force of transition electron density\cite{Beenken-J.Chem.Phys.2004-p2490}. Here we characterize the influence of the bath on the excitonic system by an Ohmic spectral density function $J(\omega)=\frac{1}{2}\pi\xi\omega e^{-\omega/\omega_c}$, where $\xi$ and $\omega_c$ is the Kondo parameter and the cutoff frequency, respectively, and their product $\xi\omega_c$ gives the reorganization energy $\lambda$. For the initial state of the exciton, we assume a normalized Gaussian wavepacket to mimic the laser induced hot exciton:
\be
\label{initial-WP}
f(E_n)=\frac{1}{Z}e^{-\frac{(E_n-\overline{E}_0)^2}{2\Delta_{E}}},
\ee
where $E_n$ is the energy of the nth eigenstate of $H_{ex}$, $\overline{E}_0$ and $\Delta_{E}$ represent the average energy and the magnitude of the energy delocalization of the initial exciton, corresponding to the pulse energy and pulse duration, respectively, $Z^2=\sum_{n}e^{-\frac{(E_n-\overline{E}_0)^2}{\Delta_{E}}}$ is the normalization constant. From \Eq{initial-WP}, we can obtain the initial wavefunction in the site representation by a simple unitary transformation, as is shown in \Fig{fig1-he}.  Several experiments indicate that the coherence length of the initial exciton may vary from $10^2$ to $10^4$\cite{Dubin-Nat.Phys.2006-p32,Chan-Nat.Chem.2012-p840,
Wuerthner-Angew.Chem.Int.Ed.2011-p3376,Whaley-J.Phys.Chem.C2014-p27235}, depending on the system of interest. Thus it is reasonable to assume an initial wavepacket with a coherence length of 100 sites in the simulation. Specifically speaking, we consider an organic aggregate chain with 200 sites, and choose the 100 sites in the center of the chain to form initial excitonic wavepacket. We set $\epsilon_n=0$ and only consider the nearest-neighbouring coupling, $V_{i,i\pm1}=1$ for simplicity.

\begin{figure}[H]
	\centering
	\includegraphics[width=0.5\textwidth]{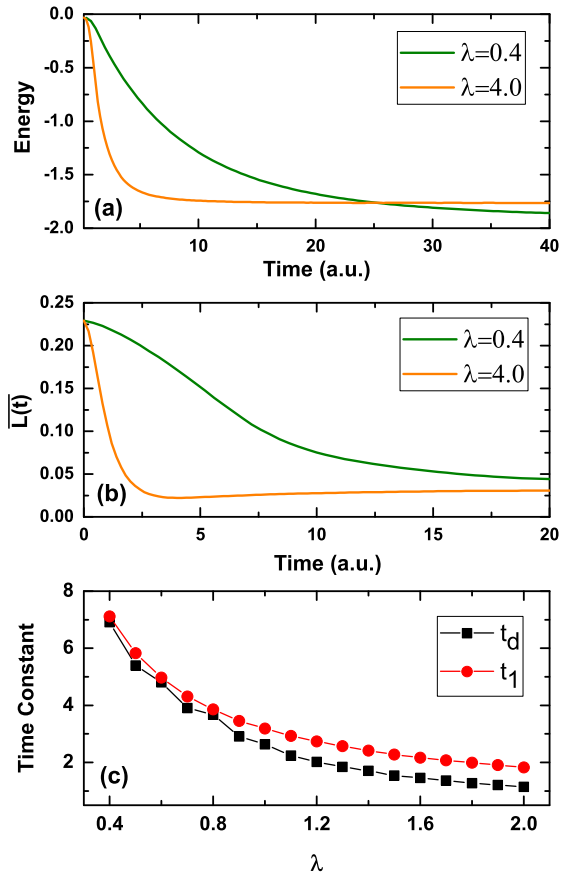}
	\caption{(a) The energy relaxation dynamics corresponding to two different exciton-phonon interaction (weak: $\lambda=0.4$ and strong: $\lambda=4.0$, respectively). (b) The  dephasing process of the averaged coherent length $\overline{L(t)}$. (c) The dephasing time ($t_d$) and the fast energy relaxation time ($t_1$) with respect to the varying reorganization energies. The initial wavepacket is generated by a laser pulse with the parameters  $\overline{E}_0=0$ ($2V_{i,i\pm 1}$ above the bottom of the energy band), $\Delta_E=0.008$, and $\omega_c=1$, $ \beta=10$. Adapted with permission from \rcite{Ke-J.Phys.Chem.Lett.2015-p1741}. Copyright 2015 American Chemical Society.}

\label{fig2-he}
\end{figure}

When there is no coupling between the exciton and the bath, the dynamics is purely coherent, exhibiting an  interference pattern of the exciton population, and the average energy of the exciton, defined as $\overline{E}(t)=\trace{\{\hat{\rho}_{ex}(t)\hat{H}_{ex}\}}$ ($\hat{\rho}_{ex}(t)$ is the reduced density operator of the exciton, and the energy relaxation curves of the hot exciton under different exciton-phonon interaction are shown in \Fig{fig2-he} (a)), will keep constant in this situation. However, the interference pattern quickly disappears when the exciton-phonon interaction is turned on, and a larger $\lambda$ leads to a faster localization of the exciton. An interesting feature is that now the energy relaxation shows a decay trend with multiple time scales, which has been observed by many experiments\cite{Banerji-J.Phys.Chem.C2011-p9726,Banerji-J.Am.Chem.Soc.2010-p17459,
Dykstra-J.Phys.Chem.B2009-p656,Freiberg-J.Phys.Chem.B1997-p7211,Chen-J.Am.Chem.Soc.2013-p18502,
Wells-J.Phys.Chem.C2007-p15404}. A possible two-step mechanism is proposed to understand this phenomenon, an initial excitonic state with energy $E_h$ first relaxes to an intermediate state with energy $E_d$ through the fast depasing dynamics, and then relaxes to the quasi-thermal equilibrium state with energy $E_e$ by slow hopping. This mechanism can be expressed as\cite{Ke-J.Phys.Chem.Lett.2015-p1741}

\be
\overline{E}(t)=E_e+Ae^{-k_1t}+Be^{-k_2t},
\ee
where $A=\frac{((E_d-E_h)k_1+(E_h-E_e)k_2)C}{k_2-k_1}$, $B=\frac{(E_e-E_d)k_1C}{k_2-k_1}$, $C$ is the initial population of the exciton, $k_1$ and $k_2$ are the rates of fast dephasing and slow hopping respectively. One thing should be noted is that when the ratio $k_1/k_2$ is larger than $\frac{E_e-E_h}{E_d-E_h}$, $A$ becomes negative, which has also been observed in experiments\cite{Banerji-J.Phys.Chem.C2011-p9726,Banerji-J.Am.Chem.Soc.2010-p17459}.

To validate the above model, we further investigate the decoherence dynamics of the exciton by calculating the coherence-length sequence$L_k(t)$, defined as
\be
L_k(t)=\sum_{i=1}^{N+1-k}|\la i|\hat{\rho}_{ex}|i+k-1\ra|,
\ee
where $|i\ra$ is a brief notation representing the localized exciton state at the ith site, and $k$ varies from $1$ to $N$, where $N$ is the number of sites. $L_k(t)$ actually reflects the coherence between two sites separated from $k-1$ sites. We calculate $L_k(t)$ for two different $\lambda$  and the results are shown in \Fig{fig2-he}(b). As can be seen, the dephasing is faster with a larger $\lambda$, consistent with experiments\cite{Wang2014dynamical}. The quantitative dephasing time $t_d$ can be obtained from the averaged coherent length $\overline{L}(t)=\sum_{k=1}^{N}L_k(t)/(N-k+1)$ by fitting $\overline{L}(t)$ exponentially. \Fig{fig2-he}(c) shows $\overline{L}(t)$ for the two different $\lambda$ cases, and one can find that $t_d$ is very close to the fast relaxation time $t_1=1/k_1$. To further identify their relationship, we calculate $t_d$ and $t_1$ for a wide range of $\lambda$, and the results are shown in \Fig{fig2-he} (c). It can be seen that these two time constants are indeed consistent with each other, validating the assumption that the fast energy relaxation is corresponding to the dephasing of the exciton.

\begin{figure}[H]
	\centering
	\includegraphics[width=0.5\textwidth]{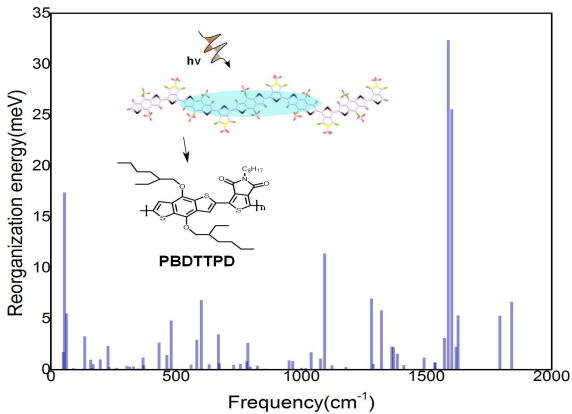}
	\caption{The frequency dependence of mode-specific reorganization energy for PBDTTPD (frequencies between 3000$cm^{-1}$ and 3300$cm^{-1}$ are not shown because the peaks are too weak to display). The molecular structure of PBDTTPD and it's component unit is embedded. Reprinted with permission from \rcite{Ke-J.Phys.Chem.Lett.2015-p1741}. Copyright 2015 American Chemical Society.}
	\label{fig6-he}
\end{figure}

As an illustrative example of realistic materials, we calculate the ultrafast exciton dynamics in a one dimensional chain of PBDTTPD copolymer, which is a promising electron donor in organic solar cells\cite{Paraecattil-J.Phys.Chem.Lett.2012-p2952,Paraecattil-J.Am.Chem.Soc.2014-p1472,Guo-J.Am.Chem.Soc.2014-p10024,
Hwang-Chem.Sci.2012-p2270,Risko-Chem.Sci.2011-p1200,Zhang-Chem.Mater.2010-p2696,Zou-J.Am.Chem.Soc.2010-p5330,VandenBrande-RSCAdv.2014-p52658}. The structure of the PBDTTPD copolymer and one of its unit are shown in \Fig{fig6-he}, the latter is regarded as a site in the Frenkel exciton model. All the parameters needed for the quantum dynamics calculations are obtained from {\it ab initio} calculations, based on the Gaussian09 and Q-chem program packages. Quantum chemistry calculations are performed on a four-unit oligomer, and alkyl sides are replaced by methyl groups to lower the computational cost. The semiempirical quantum chemical PM6 method is used to optimize the structure of the oligomer. We consider the excitation energy transfer as a reaction of $DD^\ast\rightarrow D^\ast D$. The total reorganization energy is calculated via the vibrational mode method\cite{Zhang-J.Chem.Phys.2010-p24501} as the sum of the reorganization energy from the ground state $D$ to the excited state $D^\ast$ and that from $D^\ast$ to $D$, and can be expressed as $\lambda=\sum_k\lambda_k=\sum_k\frac{1}{2}\omega_k^2\Delta Q_k^2$, where $\omega_k$ is the the frequency corresponding to the kth normal mode, $\Delta Q_k$ represents the coordinate shift between the optimized geometries of $D$ and $D^\ast$ along the kth mode. Based on the first-order approximation in the single excitation theory, the excitonic coupling can be calculated as\cite{Pan-J.Phys.Chem.C2009-p14581}
\be
\begin{split}
V_{nm}=&\la M_n^\ast M_m|\hat{H}|M_n M_m^\ast\ra \\
\approx&\int\mathrm{d}\vec{r}\int\mathrm{d}\vec{r'}\rho_n^i(\vec{r})\left[\frac{1}{|\vec{r}-\vec{r'}|}+g_{xc}(\vec{r},\vec{r'})\right]\rho_m^j(\vec{r'})-\omega_0\int\mathrm{d}\vec{r}\rho_n^j(\vec{r})\rho_m^j(\vec{r}),
\end{split}
\ee
where $\rho_n^j(\vec{r})$ is the transition density of the jth excited state of the molecule m, $\omega_0$ is the transition frequency, and $g_{xc}(\vec{r},\vec{r'})$ is the exchange-correlation potential. All other calculations, like the first excited states, are based on the long-range corrected CAM-B3LYP functional and $6-31G^\ast$ basis set from Gaussian09 package.

\begin{figure}[H]
	\centering
	\includegraphics[width=0.5\textwidth]{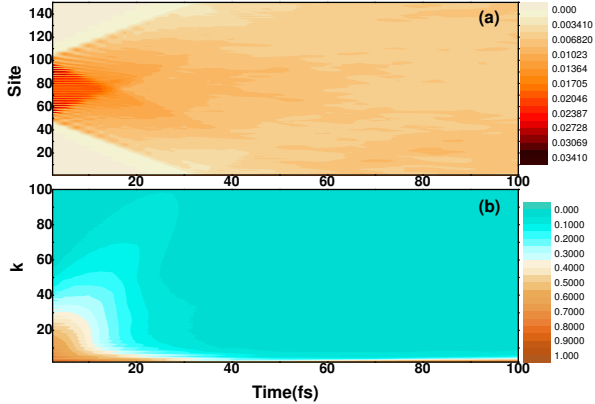}
	\caption{(a) The time-evolving exciton population of PBDTTPD at room temperature. (b) The corresponding contour map of the coherence length sequence $\{L_k(t)\}$. Reprinted with permission from \rcite{Ke-J.Phys.Chem.Lett.2015-p1741}. Copyright 2015 American Chemical Society.}
	\label{fig7-he}
\end{figure}

The mode-specific reorganization energies of PBDTTPD is shown in \Fig{fig6-he}. In the quantum dynamics simulation, an ultrafast laser pulse with a duration of 100 fs induces an initial exciton wavepacket with excess energy about 1.056 eV higher than the bottom of the exciton band. At room temperature (298 K), after 1ps evolution we obtain the two time scales of the energy relaxation, 36 fs and 491 fs, respectively. The contours of the population and coherence length sequence evolution are shown in \Fig{fig7-he}, from where one can find that the fast time (36 fs) is consistent with the dephasing process, while the slow one (491 fs) corresponds to the hopping motion, although partial coherence is still maintained. Using the femtosecond-resolved fluorescence up-conversion technique coupled with global analysis, Banerji and coworkers\cite{Banerji-J.Am.Chem.Soc.2010-p17459,Paraecattil-J.Phys.Chem.Lett.2012-p2952,Paraecattil-J.Am.Chem.Soc.2014-p1472} have shown that more than 90\% of the Stoke shift takes place within the instrumental time resolution (200 fs), and a characteristic time of about 500 fs was attributed to a single adjacent hopping (0.5-1 ps). Our simulation results are consistent with the experiments, thus we suggest that the fast time scale observed experimentally is essentially the dephasing time of exciton, whereas the slow one is corresponding to hopping motion.

\subsubsection*{\sffamily \normalsize Quantum Interferences in SF}
Much attention has been given to the SF process recently\cite{Thorsmolle2009morphology,Johnson2010high,Wang2014dynamical,Wilson2011ultrafast,
Walker2013singlet,Wilson2013temperature,Eaton2013singlet,Kolata2014molecular,Margulies2016enabling,
Bardeen2014structure,Musser2015evidence,Bakulin2016real,Zhang2016excessive} for the potential of overcoming the Shockley-Queisser limit in single-junction photovoltaic devices\cite{Shockley1961detailed,Hanna2006solar,Tayebjee2012thermodynamic}. Several works have found that the quantum interference between different pathways of the SF process, such as the direct and indirect CT-mediated pathways, can have strong impact on SF rates\cite{Tao2014electronically,Mirjani2014theoretical,Damrauer2015symmetry,Castellanos2017enhancing}.
Interestingly, there is an analogous process in aggregate spectra\cite{Kasha1963energy,Kasha1965exciton,Spano2014h,Hestand2017molecular}, where a simple dimer model are denoted as H (J)-aggregates when the direction of transition dipole moments of monomer are perpendicular (parallel) to the line connecting the the centers of monomer, with the higher (lower) excitonic states optically bright as revealed in the molecular spectra. Note that H-aggregates usually have a positive exciton-exciton coupling strength, while the J-aggregates have a negative one. Very recently, Zang and coworkers\cite{Zang2017quantum} have investigated the quantum interferences among multiple pathways in SF dynamics and found that they are closely related to the properties of J- and H-aggregates.

\begin{figure}[H]
	\centering
	\includegraphics[width=0.3\textwidth]{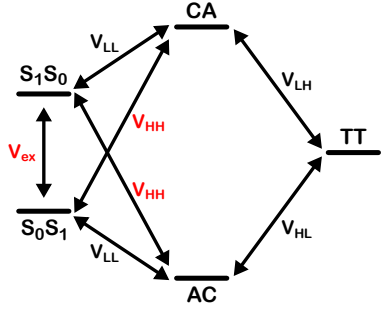}
	\caption{A schematic diagram for the five electronic states and the couplings among them. The signs of couplings marked with red are alterable, whereas the couplings marked with black are fixed to be positive. Reprinted with permission from \rcite{Zang2017quantum}. Copyright 2017 American Chemical Society.}
	\label{fig1-sf}
\end{figure}

We begin with a heterodimer model AB consisting of five electronic states denoted as $A^\ast B$ ($AB^\ast$), $A^+B^-$ ($A^-B^+$) and $A^TB^T$, corresponding to two localized singlet excited (SE) states, two CT states and a spin-correlated TT state respectively, where $A^* (B^*)$, $A^+ (B^+)$, and $A^- (B^-)$ represent the excited, positively charged, and negatively charged states of chromophore $A$ ($B$), respectively. For convenience, in the following we will refer to them as $S_1S_0$ ($S_0S_1$), CA (AC) and TT states, respectively. The electronic Hamiltonian can be expressed as
\begin{equation}
\label{H5}
\hat{H}_e =
\left(
\begin{array}{ccccc}
  E_{S_1S_0}  &   V_{ex}        &   V_{LL}  &   V_{HH}  &      0           \\
      V_{ex}     &  E_{S_0S_1}  &   V_{HH}  &   V_{LL}  &      0           \\
      V_{LL}     &  V_{HH}       &     E_{CA}   &      0          &  V_{LH}    \\
      V_{HH}    &  V_{LL}        &       0          &    E_{AC}   &  V_{HL}    \\
           0           &      0              &   V_{LH}   &   V_{HL}    &  E_{TT}     \\
\end{array}
\right).
\end{equation}
\Fig{fig1-sf} presents the relationship and the couplings among those electronic states.

By changing the signs of the five couplings $V_{ex}$, $V_{LL}$, $V_{HH}$, $V_{LH}$ and $V_{HL}$, various different quantum interference effects may appear. However, it has been shown\cite{Zang2017quantum} that the results can be sorted into several different groups. Under the assumption that the energies of the two CT states are much higher than other states, the Hamiltonian \Eq{H5} can be simplified as an effective three-levels Hamiltonian via a partitioning technique\cite{Larsson1981electron,Nitzan2006chemical}:
\begin{equation}
\label{H3}
\left(
\begin{array}{ccc}
  \widetilde{E}_{S_1S_0}  &     \widetilde{J}  &   \widetilde{V}_1     \\
    \widetilde{J}    &   \widetilde{E}_{S_0S_1} &   \widetilde{V}_2     \\
  \widetilde{V}_1  &  \widetilde{V}_2  &  \widetilde{E}_{TT}     \\
\end{array}
\right),
\end{equation}
where the effective energies are
\begin{eqnarray}
\nonumber&\widetilde{E}_{S_1S_0}=E_{S_1S_0}+\frac{V^2_{LL}}{E_{S_1S_0}-E_{CA}}+\frac{V^2_{HH}}{E_{S_1S_0}-E_{AC}},& \\
\nonumber&\widetilde{E}_{S_0S_1}=E_{S_0S_1}+\frac{V^2_{HH}}{E_{S_0S_1}-E_{CA}}+\frac{V^2_{LL}}{E_{S_0S_1}-E_{AC}},& \\
&\widetilde{E}_{TT}=E_{TT}+\frac{V^2_{LH}}{E_{TT}-E_{CA}}+\frac{V^2_{HL}}{E_{TT}-E_{AC}},&
\end{eqnarray}
and the effective couplings are
\begin{eqnarray}
\label{eff_c}
\nonumber&\widetilde{J}=V_{ex}+\frac{2 V_{LL}V_{HH}}{E_{S_1S_0}+E_{S_0S_1}-2E_{CA}}+
\frac{2 V_{HH}V_{LL}}{E_{S_1S_0}+E_{S_0S_1}-2E_{AC}},& \\
\nonumber&\widetilde{V}_1 =  \frac{2V_{LL}V_{LH}}{E_{S_1S_0}+E_{TT}-2E_{CA}} +
 \frac{2V_{HH}V_{HL}}{E_{S_1S_0}+E_{TT}-2E_{AC}},& \\
&\widetilde{V}_2 = \frac{2 V_{HH}V_{LH}}{E_{S_0S_1}+E_{TT}-2E_{CA}}+
\frac{2 V_{LL}V_{HL}}{E_{S_0S_1}+E_{TT}-2E_{AC}}.&
\end{eqnarray}
$\widetilde{V}_1$ ($\widetilde{V}_2$) represents the effective coupling between $S_1S_0$ ($S_0S_1$) state and TT state. The expression of $\widetilde{V}_1$ includes two terms corresponding to two pathways $S_1S_0$$\to$CA$\to$TT and $S_1S_0$$\to$AC$\to$TT respectively, and when terms $V_{LL}V_{LH}$ and $V_{HH}V_{HL}$ have the same (opposite) signs, the interference between the two pathways is constructive (destructive).

\begin{figure}[H]
	\centering
	\includegraphics[width=0.4\textwidth]{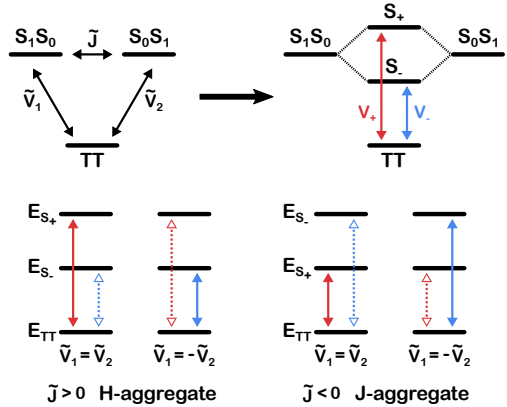}
	\caption{Schematic SF processes in J- and H-aggregates in terms of Eq. (\ref{H3-diag}). Reprinted with permission from \rcite{Zang2017quantum}. Copyright 2017 American Chemical Society.}
	\label{fig3-sf}
\end{figure}

In the following, we will investigate what effects the effective exciton-exciton coupling $\widetilde{J}$ will bring in. In this situation, an interesting analogy between the SF dynamics and the aggregate emission process\cite{Kasha1963energy,Kasha1965exciton,Spano2014h} can be drawn if we regard the TT state as the ground state and regard $\widetilde{V}_1$ and $\widetilde{V}_1$ as the transition dipoles of the two monomers. Therefore, to explain the interference effects in SF dynamics, we can directly borrow the concepts of J- and H-type aggregates by partially diagonalize the effective three-levels Hamiltonian \Eq{H5} (see the upper panel of \Fig{fig3-sf} for the schematic process):
\begin{equation}
\left(
\begin{array}{ccc}
  \widetilde{E}_{S_1S_0}  &     \widetilde{J}  &   \widetilde{V}_1     \\
    \widetilde{J}    &   \widetilde{E}_{S_0S_1} &   \widetilde{V}_2     \\
  \widetilde{V}_1  &  \widetilde{V}_2  &  \widetilde{E}_{TT}     \\
\end{array}
\right)
\to
\left(
\begin{array}{ccc}
  E_{S_+} &     0   &   V_+     \\
    0   &   E_{S_-} &   V_-     \\
   V_+    &   V_-   &   \widetilde{E}_{TT}     \\
\end{array}
\right),
\label{H3-diag}
\end{equation}
where
$E_{S_+} =\widetilde{E}_{S_1S_0}\mathrm{sin}^2\theta+\widetilde{E}_{S_0S_1}\mathrm{cos}^2\theta+
\widetilde{J}\mathrm{sin}2\theta$,
$E_{S_-} =\widetilde{E}_{S_1S_0}\mathrm{cos}^2\theta+\widetilde{E}_{S_0S_1}\mathrm{sin}^2\theta-
\widetilde{J}\mathrm{sin}2\theta $, and the two new introduced couplings are
$V_+ =\widetilde{V}_1\mathrm{sin}\theta+  \widetilde{V}_2\mathrm{cos}\theta $ and
$V_- =\widetilde{V}_1\mathrm{cos}\theta-\widetilde{V}_2\mathrm{sin}\theta $, with
$\mathrm{tan}2\theta=\frac{2\widetilde{J}}{\widetilde{E}_{S_0S_1}-\widetilde{E}_{S_1S_0}}$.

It is noted that through artificial design strategies such as changing the symmetry of covalent dimer\cite{Damrauer2015symmetry}, breaking local crystal structures\cite{Petelenz2016locally}, controlling side chains or elements\cite{Mauck2016singlet}, one can effectively tune the couplings $\widetilde{V}_1$ and $\widetilde{V}_2$. The consequence of J- and H-aggregates in aggregate spectra can be further borrowed to guide the tuning of $\widetilde{V}_1$ and $\widetilde{V}_2$. Under special assumptions that $\widetilde{E}_{S_1S_0}=\widetilde{E}_{S_0S_1}=E$, one obtains $E_{S_{\pm}} = E \pm \widetilde{J}$ and $|V_\pm| = \frac{1}{\sqrt{2}}|\widetilde{V}_1\pm \widetilde{V}_2|$. If the absolute value of $\widetilde{V}_1$ and $\widetilde{V}_2$ are further assumed to be identical, one can conclude that no matter the type of the aggregates, in the cases that $\widetilde{J}\widetilde{V}_1\widetilde{V}_2<0$, the low-energy pathway is favored and the SF rate is enhanced, while in the cases that $\widetilde{J}\widetilde{V}_1\widetilde{V}_2>0$, only the high-energy pathway is opened and the SF rate is suppressed. A schematic diagram for the above analysis is shown in the lower panel of \Fig{fig3-sf}.

\begin{figure}[H]
	\centering
	\includegraphics[width=0.5\textwidth]{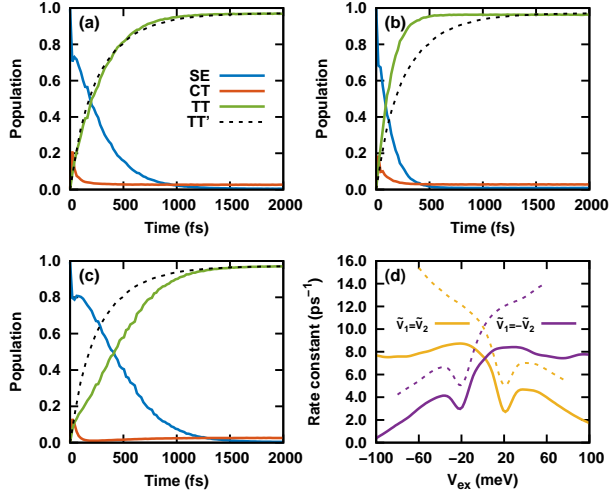}
	\caption{Population evolution of SF with (a) $V_{ex}=20$ meV, (b) $V_{ex}=-40$ meV and (c) $V_{ex}=90$ meV, as well as TT state population from only one SE state pathway (dashed line).
(d) The SF rates in terms of $V_{ex}$ in dimer (solid line) and ten monomer (dashed line) models. Reprinted with permission from \rcite{Zang2017quantum}. Copyright 2017 American Chemical Society.}
	\label{fig4-sf}
\end{figure}

This effective three-levels picture is only valid when the energies of CT states are much higher than the energies of other electronic states. Beyond this regime, it is not easy to find the interference effects analytically. Besides, The complex electron-phonon interaction should also be incorporated in the analysis. In the following the perturbative non-Markovian stochastic Schr\"odinger equation is adopted to investigate the interference effects behind J- and H-aggregates, where the Debye spectral density \Eq{SD-DB} is adopted for all the five electronic states with an identical set of parameters, $\lambda=50$ meV and $\omega_c=1450$ cm$^{-1}$\cite{Berkelbach-J.Chem.Phys.2013-p114103}, and the temperature being 300K. The electronic Hamiltonian elements are set as as $E_{S_1S_0}-E_{TT}=200$ meV, $V_{LL}=V_{LH}=V_{HL}=V_{HH}=50$ meV, and $E_{CA}(E_{AC})-E_{TT}=400$ meV\cite{Greyson2010maximizing,Tao2014electronically,Chan2013quantum,Wang2014maximizing}, whereas $V_{ex}$ is a variable that controlling the type of the aggregates. The initial wavefunction of the exciton is set as $(|S_1S_0\ra+|S_0S_1\ra)/\sqrt{2}$. \Fig{fig4-sf}(a)-(c) show the population evolution with $V_{ex}=20, -40, 90$ meV, corresponding to the cases when $\widetilde{J}=0$, $<0$, and $>0$ respectively. It is clear that when $\widetilde{J}=0$, the interference between two SE states disappears\cite{Hestand2017molecular,Hestand2015interference}, therefore the SF process should be similar to that of the null-aggregate case, as can be seen in \Fig{fig4-sf}(a). It is also clear in \Fig{fig4-sf}(b)-(c) that the negative (positive) $\widetilde{J}$ speeds up (suppresses) the SF processes as expected. Via an exponential fitting procedure $P_{SE}(t)=e^{-kt}$, we can obtain the SF rate $k$, and the results in terms of $V_{ex}$ are shown in \Fig{fig4-sf}(d). Focusing on the case of $\widetilde{V}_1=\widetilde{V}_2$, it is clear that the SF rates when $V_{ex}<20$ meV are larger than those when $V_{ex}>20$ meV, totally consistent with the explanation borrowed from J- and H-aggregates. However, in the region of $V_{ex}>20$ meV, an interesting behavior is that the SF rates first increase and then decrease with the increase of $V_{ex}$. In this region, the pathway from the lower-energy state $S_-$ (high-efficient one) to the TT state is prohibited, while that from the higher-energy state $S_+$ (low-efficient one) is allowed. When $\widetilde{J}$ is small, the energy splitting between $S_-$ and $S_+$ states is not so high that the low-efficient pathway may still enhance the SF rates as compared to the null aggregate case, since $\widetilde{V}_+$ is larger than $\widetilde{V}_1$ and $\widetilde{V}_2$. When $\widetilde{J}$ is high enough, the transition from $S_+$ state to TT state becomes inefficient due to the large energy gap, therefore the SF rates decrease.

In the above analysis, the J-aggregates with negative $\widetilde{J}$ allow both the non-radiative SF and radiative emission pathways, which is not favored from a practical view since the two processes may have similar time scales\cite{Yost2014transferable}. One can suppress one of the pathways by controlling the values of $\widetilde{V}_1$ and $\widetilde{V}_2$. For example, let $V_{LL}\to -V_{LL}$ and $V_{LH}\to -V_{LH}$, we can get $\widetilde{V}_1=-\widetilde{V}_2$ (see \Eq{eff_c}), and the roles of J- and H-aggregates are totally inversed, as is seen from the purple line in \Fig{fig4-sf}(d). To show that these results are general, the same calculations are also done for a one-dimensional chain model with 10 monomers\cite{Zang-J.Phys.Chem.C2016-p13351}, and the same relationship between SF rates and $V_{ex}$ is obtained as shown in \Fig{fig4-sf}(d). The only discrepancy is that the SF rates are larger than those of the dimer model, which is due to the exciton delocalization effect\cite{Pensack2015exciton,Zang-J.Phys.Chem.C2016-p13351}.

\section*{\sffamily \Large CONCLUSIONS}
Based on the path integral formalism of an open system coupled to harmonic baths, we have outlined a group of stochastic wavefunction-based approaches, i.e., non-Markovian stochastic Schr\"odinger equations, which are presented in a progressive manner in terms of the level of approximations so as to  set up an integrated platform  for its versatile applications in various physical and chemical scenarios.
First of all, we have introduced a mixed strategy combining the stochastic processes and the hierarchical techniques to obtain a hierarchical set of stochastic differential equations, which are numerically exact and suitable for the cases where at least part of the bath correlation functions exhibit exponential decaying features. We have further demonstrated that within the Debye bath, by choosing the residual bath correlation function to be a pure imaginary function, one can obtain a specific set of equations which require the smallest number of the auxiliary wavefunctions, effectively avoiding the perplexity along with the low temperature conditions in other hierarchical quantum master equations for density matrix. Then, considering the cases where the system-bath coupling is not quite strong, the bath is highly structured, or the system size is very large, we further summarized a systematic perturbation expansion scheme with respect to the system-bath coupling. Under the lowest order approximation, the perturbative stochastic Schr\"odinger equations are found to be reliable in a very broad parameter range, which is highly commendable since it arrives at an optimal balance between accuracy and computational cost. Of course, the equations can be further simplified if the systems are highly-symmetric or the bath memory is extremely short-lived. In the last part, a few applications of the above-mentioned approaches are presented, including the calculations of  excitation energy transfer in the FMO complex, quantum transport properties of hot exciton in a copolymer chain and the studies about the quantum interference effects on the the promising SF processes in organic materials.  All in all, the non-Markovian stochastic Schr\"odinger equations due to its distinctive advantages, are expected to play a crucial role in the nanoscale realistic systems in the near future.

\section*{\sffamily \Large ACKNOWLEDGEMENTS}
This work is supported by the National Science Foundation of China (Grant No. 21573175 and 21773191).




\end{document}